# XFluids: A unified cross-architecture heterogeneous reacting flows simulation solver and its applications for multi-component shock-bubble interactions[a]

Jinlong Li[1], Shucheng Pan[1,2,3,b]

[1] *School of Aeronautics, Northwestern Polytechnical University, Xi'an, 710072, China*

[2] *Institute of Extreme Mechanics, Northwestern Polytechnical University, Xi'an 710072, China*

[3] *National Key Laboratory of Aircraft Configuration Design, Xi'an 710072, China*

We present a cross-architecture high-order heterogeneous Navier-Stokes simulation solver, XFluids, for compressible reacting multicomponent flows on different platforms. The multi-component reacting flows are ubiquitous in many scientific and engineering applications, while their numerical simulations are usually time-consuming to capture the underlying multiscale features. Although heterogeneous accelerated computing is significantly beneficial for large-scale simulations of these flows, effective utilization of various heterogeneous accelerators with different architectures and programming models in the market remains a challenge. To address this, we develop XFluids by SYstem-wide Compute Language (SYCL), to perform acceleration directly targeted to different devices, without translating any source code. A variety of optimization techniques have been proposed to increase the computational performance of XFluids, including adaptive range assignment, partial eigensystem reconstruction, hotspot device function optimizations, etc. This solver has been open-sourced, and tested on multiple graphics processing units (GPUs) from different mainstream vendors, indicating high portability. Through various benchmark cases, the accuracy of XFluids is demonstrated, with approximately no efficiency loss compared to existing GPU programming models, such as Compute Unified Device Architecture (CUDA) and Heterogeneous-computing Interface for Portability (HIP). In addition, the Message Passing Interface (MPI) library is used to extend the solver to multi-GPU platforms, with the GPU-enabled MPI supported. With this, the weak scaling of XFluids for multi-GPU devices is larger than 95% for 1024 GPUs. Finally, we simulate both the inert and reactive multicomponent shock-bubble interaction problems with high-resolution meshes, to investigate the reacting effects on the mixing, vortex stretching, and shape deformation of the bubble evolution.

## I. INTRODUCTION

Multi-component reacting compressible flows are ubiquitous in aerospace, combustion, explosion, astrophysics, and other fields. One example of these flows is the multicomponent shock-bubble interaction (SBI)[1], which is encountered in e.g. the supersonic combustion systems[2], the atmospheric sonic boom propagation[3], and the shock propagation through foams and bubbly liquids[4], etc. Compared to experiments[5], numerical simulations of multicomponent SBIs provide more physical details and are more flexible in extreme conditions. However, these flows are governed by complex physics coupling of convection, dissipation, reaction, species diffusion, and chemical reactions, and usually contain disparate scales of fluid dynamics and reactions. Consequently, numerical simulations of compressible multicomponent reacting flows require significantly larger

---

computing resources than those in non-reacting flows, especially for a large number of components, indicating an urgent need of efficient acceleration technique.

Recently, heterogeneous devices like graphics processing units (GPUs) have been widely applied to accelerating numerical simulations of many scientific problems, and show much faster computational speed than that of traditional central processing unit (CPU) computing. Different GPU hardware architectures have been introduced by multiple vendors, including Nvidia Corporation (NVIDIA), Advanced Micro Devices (AMD), Intel Corporation (Intel), Apple Inc. (Apple), etc. And the corresponding GPU programming standards have been designed to appropriately enable those devices, e.g. Compute Unified Device Architecture (CUDA), Heterogeneous-computing Interface for Portability (HIP), Open Accelerators (OpenACC), Open Computing Language (OpenCL), etc. Based on these devices and standards, numerous computational fluid dynamics (CFD) solvers have been developed, of which many are open-sourced. For example, based on CUDA, STREAmS[6], and HYPAR[7] are dedicated to the direct numerical simulation of ideal-gas turbulent flows. OpenCFD-SCU[8] solves compressible turbulence by CUDA and HIP, and achieves an approximately ideal weak-scaling efficiency for $10^4$ AMD GPUs. Similarly, Wang et al.[9] implement the high-order gas-kinetic scheme for compressible turbulence simulations based on Message Passing Interface (MPI) and CUDA. In addition, specifically designed for high-order flux reconstruction schemes, ZEFR[10] solves ideal-gas compressible viscous flows by CUDA, and PyFR[11] solves advection-diffusion equations accelerated by AMD and NVIDIA GPUs. Meanwhile, Xiang et al.[12] also utilize CUDA and HIP to accelerate lattice Boltzmann method (LBM) simulations of turbulent square duct flow at high Reynolds numbers with 1.57 billion grids and 384 GPUs. For multiphase problems, FluTAS[13] accelerates the finite difference method for incompressible multiphase flows by OpenACC, and JAX-Fluids[14] simulates compressible two-phase flows on NVIDIA GPUs and Google tensor processing units (TPUs), with machine learning automatic differentiation capabilities. Furthermore, to accelerate combustion simulations, DeepFlame[15] is developed on top of OpenFOAM[16] as its CFD framework, with reaction kinetics integrated through Cantera or neural-network models, and recently has been coupled with NVIDIA's Algebraic Multigrid Solver (AmgX) library to enable GPU computing[17]. Based on the OpenCL standard, Gorobets and Bakhvalov[18] develop a large-scale heterogeneous parallel compressible turbulence numerical simulation solver, and Moritz et al. released FluidX3D[19] as an efficient LBM solver, which both provide a cross-platform solution for the heterogeneous computations.

With such diverse devices and models, ensuring both cross-architecture and portability features of heterogeneous simulation software is challenging. The computational software developed for a specific manufacturer's GPU devices and programming ecosystem is difficult to port to others with acceptable computational efficiency. Although OpenCL shows strong cross-architecture capability, its low-level application programming interface (API) usually is cumbersome for developers, and is relatively difficult to optimize OpenCL-based CFD solvers to achieve comparable efficiency with native models (CUDA or HIP). To overcome this issue, one strategy is using an abstract middleware layer that separates the application software from hardware and its programming model, such as Trilinos[20], RAJA[21], Ginkgo[22], and Kokkos[23]. For example, KARFS[24] is developed based on Kokkos and MPI for direct numerical simulation of reacting flows. Another strategy is developing cross-architecture CFD codes based on the SYstem-wide Compute Language (SYCL)[25] programming standard, which invokes CPUs, FPGAs, and GPUs of different vendors without any translation of the source code. SYCL has now been implemented in PeleLMeX[26] and FUN3D[27] for reacting flow simulations. However, the former only focuses on low Mach number reacting flows, while the latter is an in-house code. To simulate the multicomponent SBIs phenomenon on various GPU backends by SYCL, this paper proposes XFluids, an open-source CFD tool for compressible reacting flows, with high-order accuracy and cross-architecture capability. The remaining of this paper is structured into 6 sections, each contributing to a comprehensive understanding of XFluids. First, the physical models and the methods employed by XFluids are detailed in Section II. Next, the solver's structure and GPU optimization strategies are shown in Section III, where we detail the solver architecture, emphasize the intricacies of its implementation on GPUs, and thoroughly examine the optimization strategies tailored for efficient GPU execution. The validation on various examples in Section IV demonstrates the capability and accuracy of XFluids, while the performance and scalability across multiple different GPU devices are assessed in Section V. Finally, this solver is applied for high-resolution numerical simulations of inert and reactive SBIs in Section VI, followed by a conclusion in Section VII.

## II. METHODOLOGY

### A. Governing equations

The compressible reacting multicomponent Navier-Stokes equations are solved in XFluids, i.e.,

$$\frac{\partial \mathbf{U}}{\partial t} + \nabla \cdot \mathbf{F}(\mathbf{U}) = \nabla \cdot \mathbf{F}_v(\mathbf{U}) + \mathbf{S}, \tag{1}$$

with

$$\mathbf{U} = \begin{pmatrix} \rho \\ \rho \mathbf{u} \\ E \\ \rho Y_s \end{pmatrix}, \quad \mathbf{F}(\mathbf{U}) = \begin{pmatrix} \rho \mathbf{u} \\ \rho \mathbf{u}\mathbf{u} + p\mathbf{I} \\ (E+p)\mathbf{u} \\ \rho \mathbf{u} Y_s \end{pmatrix}, \quad \mathbf{F}_v(\mathbf{U}) = \begin{pmatrix} 0 \\ \boldsymbol{\tau} \\ \boldsymbol{\tau} \cdot \mathbf{u} - \mathbf{q}_c - \mathbf{q}_d \\ \mathbf{J}_s \end{pmatrix}, \quad \text{and} \quad \mathbf{S} = \begin{pmatrix} 0 \\ 0 \\ \dot{\omega}_T \\ \dot{\omega}_s \end{pmatrix}. \tag{2}$$

Here $\rho$ represents the mixture density, $\boldsymbol{u}$ the velocity vector, $E$ the total energy, $p$ the pressure, and $Y_s$ the mass fraction of species $s = 1, 2, \ldots N-1$ with $N$ the total number of species. The viscous stress tensor $\tau$ for a Newtonian fluid is defined as

$$\tau = 2\bar{\mu}\left[\frac{1}{2}\left(\nabla\boldsymbol{u} + (\nabla\boldsymbol{u})^T\right) - \frac{1}{3}\mathbf{I}(\nabla\cdot\boldsymbol{u})\right], \tag{3}$$

and the heat conduction is defined by the Fourier law,

$$\boldsymbol{q}_c = -\bar{\kappa}\nabla T. \tag{4}$$

The interspecies diffusional heat flux $\boldsymbol{q}_d$ [28] in Eq. (2) is defined by

$$\boldsymbol{q}_d = \sum_{s=1}^{N} h_s \boldsymbol{J}_s, \tag{5}$$

where $h_s$ and $\boldsymbol{J}_s$ are the individual species enthalpy and the diffusion[29]

$$\boldsymbol{J}_s = -\rho\left(D_{sm}\cdot\nabla Y_s - Y_s \sum_{i=1}^{N} D_{jm}\cdot\nabla Y_s\right), \tag{6}$$

respectively, with $D_{sm}$ indicting the effective binary diffusion coefficient of species $s$ diffusing into the mixture, with the subscript $m$ corresponding to the mixture.

The multicomponent gas mixture follows the ideal-gas equation of state,

$$p = \rho\bar{R}T, \quad \bar{R} = R_u\sum_{s=1}^{N}\frac{Y_s}{W_s} \tag{7}$$

with $\bar{R}$ being the specific gas constant of the mixture defined by the universal gas constant $R_u$, the molar mass $W_s$, and the mass fraction $Y_s$.

**B. Reaction kinetics**

Chemical reaction kinetics are represented by the source term in Eq. (1), which contains the heat release $\dot{\omega}_T$ and species formation and destruction in terms of individual mass rates $\dot{\omega}_s$. The specific heat release is defined as

$$\dot{\omega}_T = -\sum_{s=1}^{N} \Delta h_{f,s}^0 \dot{\omega}_s, \tag{8}$$

where $\Delta h_{f,s}^0$ is the heat of formation of each species $s$. New temperature $T$ is updated by solving $\dot{\omega}_T$, and $Y_s$ is computed through solving the mass production rates $\dot{\omega}_s$,

$$\dot{\omega}_s = W_s \sum_{r=1}^{N_R} \upsilon_{s,r} \Gamma_r \left( k_{f,r} \prod_{s=1}^{N} [X_s]^{\upsilon'_{s,r}} - k_{b,r} \prod_{s=1}^{N} [X_s]^{\upsilon''_{s,r}} \right), \tag{9}$$

with $N_R$ being the number of reactions, $\Gamma_r$ the third body efficiency, $X_s$ the molar concentration of reactants and products, respectively. In addition, $\upsilon'_{s,r}$ and $\upsilon''_{s,r}$ are the molar stoichiometric coefficients of the reactant and the product for reaction $r$, respectively. Then, the net stoichiometric coefficient $\upsilon_{s,r}$ becomes

$$\upsilon_{s,r} = \upsilon''_{s,r} - \upsilon'_{s,r}. \tag{10}$$

In Eq. (9), the forward and backward reaction rates $k_{f,r}$ and $k_{b,r}$ of reaction $r$ are calculated by the Arrhenius law,

$$k_{f,r} = A_{f,r} T^{B_{f,r}} \exp\left(\frac{E_{f,r}}{R_u T}\right), \quad k_{b,r} = \frac{k_{f,r}}{K_{c,r}}, \tag{11}$$

where $A_{f,r}$ is the pre-exponential factor, $B_{f,r}$ is the temperature exponent and $E_{f,r}$ is the activation energy[30] for each reaction $r$. The equilibrium constants $K_{c,r}$ are used to estimate the backward reaction rates,

$$K_{c,r} = \left(\frac{p^\circ}{R_u T}\right)^{\sum_{s=1}^{N} \upsilon_{s,r}} \exp\left(\sum_{s=1}^{N} \upsilon_{s,r} \left(\frac{S_s}{R_s} - \frac{h_s}{R_s T}\right)\right), \tag{12}$$

where $p^\circ = 1$ atm, the entropy $S_s$ and the enthalpy $h_s$ of the species $s$ can be fitted by NASA thermodynamic polynomials[31].

## C. Numerical method

### 1. Space and time discretization

XFluids utilizes high-order finite-difference discretization schemes to approximate the spatial fluxes in Eq. (2),

$$\left(\frac{\partial \mathbf{F}}{\partial x}\right)_i = \left(\hat{\mathbf{F}}_{i+\frac{1}{2}} - \hat{\mathbf{F}}_{i-\frac{1}{2}}\right) \cdot \frac{1}{\Delta x} + \mathbf{O}\left(\Delta x^{2k-1}\right), \tag{13}$$

where $\hat{\mathbf{F}}_{i+1/2}$ is the numerically approximated convective or viscous fluxes at the cell interface, and the parameter $k$ is determined by the order of the numerical discretization schemes. For the convective fluxes, its Jacobian $A$ is decomposed and diagonalized by $\Lambda = LAR$, where $\Lambda$ is the diagonal eigenvalue matrix, $L$ and $R$ are the left and right eigenvectors, respectively. Here, the eigen system for multi-component systems is given by the existing literature[32,33]. Then, on each reconstruction stencil, the split positive and negative characteristic fluxes can be obtained by projecting the cell average flux and the conservative variables onto the eigenspace,

$$\mathbf{F}_m^\pm = \frac{1}{2} \mathbf{L}_{i+\frac{1}{2}} \cdot \left(\mathbf{F}_m \pm \widetilde{\mathbf{\Lambda}} \mathbf{U}_m\right), \tag{14}$$

where $m$ ranges from $i-k+1$ to $i+k$, $\mathbf{L}_{i+1/2}$ is the left eigenmatrix of the locally linearized Roe-averaged flux Jacobian matrix, and $\widetilde{\mathbf{\Lambda}}$ denotes the artificial viscosity coefficient. Later, those characteristic fluxes $\mathbf{F}_m^\pm$ are reconstructed to obtain the positive and negative numerical fluxes $\mathbf{F}_{i+1/2}^\pm$, which are transformed through the right eigenvectors, leading to the cell-interface flux $\hat{\mathbf{F}}_{i+1/2}$ in Eq. (13),

$$\hat{\mathbf{F}}_{i+\frac{1}{2}} = \mathbf{R}_{i+\frac{1}{2}} \cdot \left(\mathbf{F}_{i+\frac{1}{2}}^+ + \mathbf{F}_{i+\frac{1}{2}}^-\right), \quad \mathbf{F}_{i+\frac{1}{2}}^\pm = \text{REC}\left(\mathbf{F}_m^\pm\right). \tag{15}$$

Here $\text{REC}(\cdot)$ indicates the specific reconstruction scheme for the numerical fluxes, which can be the central-upwind 6$^{\text{th}}$-order weighted essentially non-oscillatory (WENO-CU6) scheme[34], the 5$^{\text{th}}$-order WENO scheme[35], and the 7$^{\text{th}}$-order WENO scheme[36] in XFluids. The multicomponent Roe-average treatment[37] is applied to avoid spurious pressure and temperature oscillations[38]. For the viscous flux, its numerical approximation is achieved by using a central 4$^{\text{th}}$-order accuracy scheme[39].

After the discretization of the fluxes, the semi-discrete equation of Eq. (1) is advanced by the 3$^{\text{rd}}$-order Strong-Stability-Preserving (SSP) Runge-Kutta time-marching scheme[40],

$$\begin{aligned} \mathbf{U}^1 &= \mathbf{U}^n + \Delta t \cdot \mathbf{L}(\mathbf{U}^n) \\ \mathbf{U}^2 &= \frac{3}{4}\mathbf{U}^n + \frac{1}{4}\left[\mathbf{U}^1 + \Delta t \cdot \mathbf{L}(\mathbf{U}^1)\right] \\ \mathbf{U}^{n+1} &= \frac{1}{3}\mathbf{U}^n + \frac{2}{3}\left[\mathbf{U}^2 + \Delta t \cdot \mathbf{L}(\mathbf{U}^2)\right] \end{aligned} \quad (16)$$

where $\mathbf{L}(\mathbf{U})$ is the residual of Eq. (1).

### *2. Positivity preserving*

To enhance the numerical stability of the solver, a positivity-preserving method[41] for high-order conservative schemes is extended to both the density $\rho$ and mass fractions $Y_s$. For instance, without loss of generality, we consider the first-order Euler time marching scheme,

$$\mathbf{U}_i^{n+1} = \mathbf{U}_i^n + \lambda \cdot \left(\mathbf{F}_{i+\frac{1}{2}} - \mathbf{F}_{i-\frac{1}{2}}\right), \quad \lambda = \Delta t / \Delta x = \frac{CFL}{\widetilde{\Lambda}}, \quad (17)$$

where the superscript $n$ and $n+1$ represent the old and new time steps, respectively. In this case, the Eq.(17) can be re-formed as a convex combination,

$$\mathbf{U}_i^{n+1} = \frac{1}{2}\left(\mathbf{U}_i^n + 2\lambda \mathbf{F}_{i-\frac{1}{2}}\right) + \frac{1}{2}\left(\mathbf{U}_i^n - 2\lambda \mathbf{F}_{i+\frac{1}{2}}\right) = \frac{1}{2}\mathbf{U}_i^- + \frac{1}{2}\mathbf{U}_i^+. \quad (18)$$

As the first-order Lax-Friedrichs flux,

$$\mathbf{F}_{i+\frac{1}{2}}^{LF} = \frac{1}{2}\left[\mathbf{F}_i + \mathbf{F}_{i+1} + \widetilde{\Lambda} \cdot \left(\mathbf{U}_i^n - \mathbf{U}_{i+1}^n\right)\right], \quad (19)$$

is inherently positivity preserving under a condition of $CFL \leq 0.5$, we can blend the Lax-Friedrichs flux and the high-order numerical flux $\hat{\mathbf{F}}_{i+1/2}$ in Eq.(13) by using a limiting factor $0 \leq \theta \leq 1$,

$$\mathbf{F}_{i+\frac{1}{2}}^* = (1-\theta)\mathbf{F}_{i+\frac{1}{2}}^{LF} + \theta \mathbf{F}_{i+\frac{1}{2}}, \quad (20)$$

see the original publication[41] for calculating the limiting factor. Finally, the blended flux is used to update Eq. (13).

### *3. Calculation of the thermodynamics and transport properties*

After solving Eq.(13) by the prescribed space and time discretization schemes above, the conservative states $U$ is updated, which can be used to obtain the thermodynamic states, e.g. the density, pressure, and temperature, etc. For instance, the pressure can be solved by the ideal-gas EOS in Eq.(7), where the temperature $T$ is updated by Newton's iteration method as

$$T^{k+1} = T^k + \frac{e - e(T^k)}{\partial e / \partial T} = T^k + \frac{e - e(T^k)}{C_v(T^k)}$$

$$e(T^k) = \sum_{i=1}^{N} Y_s e_s(T^k) = \sum_{i=1}^{N} Y_s h_s(T^k) - \overline{R} T^k \qquad (21)$$

$$C_v(T^k) = \sum_{i=1}^{N} Y_s C_{v,s}(T^k) = \sum_{i=1}^{N} Y_s C_{p,s}(T^k) - \overline{R}$$

Here $e$ is the internal energy of the mixture, $C_v$ is the constant-pressure specific heat capacity of the mixture. $C_{p,s}$ and $h_s$ are fitted by the NASA Glenn thermodynamics polynomials[31].

There are three kinds of coefficients in the gas transport model of XFluids, i.e. viscosity $\mu_s$ and thermal conductivity $\kappa_s$, binary diffusion coefficients $D_{st}$, with the subscript $t$ corresponding to the species $t$. Here, both the mixture-averaged viscosity $\bar{\mu}$ and thermal conductivity $\bar{\kappa}$ are given by the semiempirical formula[42,43]

$$\bar{\alpha} = \sum_{s=1}^{N} \frac{X_s \alpha_s}{\sum_{t=1}^{N} X_t \Phi_{st}^{\alpha}}, \quad \Phi_{st}^{\alpha} = \frac{1}{\sqrt{8}} \left(1 + \frac{W_s}{W_t}\right)^{-\frac{1}{2}} \left(1 + \left(\frac{\alpha_s}{\alpha_t}\right)^{\frac{1}{2}} \left(\frac{W_t}{W_s}\right)^{\frac{1}{4}}\right)^2, \qquad (22)$$

with the $\bar{\alpha}$ as $\bar{\mu}$ or $\bar{\kappa}$, and $\alpha_s$ as the pure gas viscosity $\mu_s$ or the thermal conductivity $\kappa_s$, $W_s$ and $X_s$ denote the molecular weight and the molecular concentration fraction of species $s$, respectively. $\mu_s$ is calculated from the standard kinetic theory[44]

$$\mu_s = \frac{5}{16} \frac{\sqrt{\pi W_s k_B T}}{\pi \sigma_s^2 \Omega^{(2,2)^*}}, \qquad (23)$$

with $\pi$, $k_B$, $\sigma_s$ as the circular constant, the Boltzmann's constant, and the collision diameter, respectively. $\Omega^{(2,2)*}$ is the second category collision integral for viscosity, and $\kappa_s$ is assumed to be composed of translational, rotational, and vibrational contributions as given by Warnatz[45],

$$\kappa_s = \frac{\eta_s}{W_s}\left(f_{trans}C_{\upsilon,trans} + f_{rot}C_{\upsilon,rot} + f_{vib}C_{\upsilon,vib}\right). \tag{24}$$

$D_{st}$ is defined by the reduced collision diameter $\sigma_{st}$, the first category collision integral for diffusion $\Omega^{(1,1)*}$, and the reduced molecular mass for the $(s,t)$ species pair $W_{st}$.

$$D_{st} = \frac{3}{16}\frac{\sqrt{2\pi k_B^3 T^3/W_{st}}}{p\pi\sigma_{st}^2\Omega^{(1,1)*}}, \quad W_{st} = \frac{W_s W_t}{W_s + W_t} \tag{25}$$

The two collision integral $\Omega^{(1,1)*}$ and $\Omega^{(2,2)*}$ are interpolated by the published tables[46]. Finally, $D_{sm}$ in Eq.(6) is computed from the constitutive empirical law[43],

$$D_{sm} = \frac{\sum_{t=1,t\neq s}^{N} X_t W_t}{\overline{W}\sum_{t=1,t\neq s}^{N} X_t/D_{st}}, \quad \overline{W} = \sum_{s=1}^{N} X_s W_s, \tag{26}$$

where $\overline{W}$ is the molecular mass of the mixture.

### 4. Reaction integral

The first-order Lie-Trotter time splitting scheme[47] and the second-order Strang time splitting scheme[48] are both available in XFluids to decompose the stiff chemical reaction term from the Navier-Stokes equations Eq.(1). The resulting system of the partial differential equations (PDE) and the stiff system of ordinary differential equations (ODE) is solved sequentially. In particular, the stiff ODE system is solved by a Quasi-Steady-State (α-QSS) solver CHEMEQ2[49] which is efficient and accurate as well as easy to be coupled with existing multicomponent flow simulation models[50], the reaction kinetics Eq.(9) can be expressed as well as follow,

$$\frac{dY_s}{dt} = g_s = q_s - p_s Y_s, \tag{27}$$

where the source term of the species $s$ can be written as the difference of the production rate $q_s$ and the loss rate $p_s Y_s$. To numerically solve Eq.(27) over the total time duration, a discrete-time integration scheme with multiple time steps is adopted. At each time step $\Delta t$, a multi-step integration method of prediction-correction is applied, which consists of a predictor step that estimates the solution at the next time level and multiple corrector steps that improve the accuracy of the solution by using the predicted values,

$$\begin{aligned}
Y_s^p &= Y_s^0 + \frac{\Delta t \left( q_s^0 - p_s^0 Y_s^0 \right)}{1 + \alpha_s^0 p_s^0 \Delta t} \\
Y_s^c &= Y_s^0 + \frac{\Delta t \left( q_l - \bar{p}_l Y_s^0 \right)}{1 + \bar{\alpha}_l \bar{p}_l \Delta t} \\
\alpha_s(p_s \Delta t) &= \frac{1 - \left(1 - e^{-p_s \Delta t}\right)}{\left(1 - e^{-p_s \Delta t}\right) p_s \Delta t} \\
\bar{p}_l &= \frac{1}{2}\left( p_s^0 + p_s^p \right), \quad q_l = \bar{\alpha}_l q_s^p + (1 - \bar{\alpha}_l) q_s^0
\end{aligned} \tag{28}$$

The existing mechanisms usually show large discrepancies in the number of reactions and species, as well as the consideration of third-body efficiencies, and duplicated and pressure-dependent reactions. Therefore, a complex mechanism, such as $H_2$-$O_2$ combustion consisting of 9 species involving 18 reversible reactions[51] or 19 reactions[52], 21 reactions[53], is necessary to predict the correct ignition delay time.

## III. IMPLEMENTATION DETAILS

### A. Parallel execution

To reveal the process of SYCL parallelism, a complete heterogeneous parallel task processing program sample performing vector summation based on SYCL is presented in **APPENDIX A**. In general, XFluids follows the SYCL 2020 Specification to achieve parallel computing. Its parallel task consists of the following 5 steps: (1) execute *"sycl-ls"* or *"acpp-info"* to acquire all available devices before running the code; (2) then a queue *"q"* is initialized and associated with the selected device, where the pointer-based Unified Shared Memory (USM) model is used for easier managing attached memories of accelerator device; (3) next, choose appropriate ND-Range configuration, which is an SYCL implementation of the thread hierarchy; (4) and the parallelism submission and alternative reduction operation are intrinsically tied to the ND-Range assignment; (5) finally,

*"q.wait()"* serves as a synchronization point between the host and the device, ensuring that data dependencies are maintained. SYCL code can optionally be compiled by various compilation systems such as Intel oneAPI[54] and AdaptiveCpp[55]. Intel oneAPI introduced in 2019 fully utilizes CPUs, FPGAs, and Intel GPUs, with optional Codeplay's plugins[56] enabling support for AMD and NVIDIA platforms. AdaptiveCpp usually aggregates existing clang toolchains and augments them with support for SYCL constructs. This means that the AdaptiveCpp compiler can not only compile SYCL code but also CUDA/HIP code even if they are mixed in the same source file, making all CUDA/HIP features and even vendor-optimized libraries such as rocPRIM or CUB also available in SYCL code.

### 1. Device selection

Upon the successful installation of all necessary dependencies, the utilization of the *"sycl-ls"* command-line tool, provided by oneAPI, becomes instrumental in enumerating the available device backends. The standard output from this utility typically encompasses a CPU backend alongside a CPU-emulated FPGA backend and multiple intel-GPU backends. For non-intel GPUs, the presence of Codeplay's plugin can append the GPU backends of other vendors (e.g. NVIDIA or AMD) to the enumerated devices, thereby enriching the device landscape accessible to the developer. In a similar manner, the execution of the *"acpp-info"* tool, which is a component of AdaptiveCpp, yields comprehensive details of the available devices, along with their respective capabilities. This information is crucial for developers to make informed decisions regarding device selection and optimization strategies. FIG. **1** exemplifies the output generated by both *"sycl-ls"* and *"acpp-info"* on a desktop with an 8-Core CPU and a NVIDIA RTX 3070 GPU, and a node of heterogeneous supercomputer appended with a 32-Core CPU and 4 AMD Instinct MI50. With this, developers and users can tailor their applications to leverage the unique strengths of each device, leading to optimized performance and enhanced application capability.

```
$ acpp-info
================Backend information==================
Loadedbackend 0: OpenMP
Founddevice: hipSYCL OpenMP host device
Loadedbackend1: CUDA
Founddevice: NVIDIA GeForce RTX 3070

$ sycl-ls
================Backend information==================
[opencl:acc:0] Intel(R) FPGA Emulation Platform for OpenCL(TM), Intel(R) FPGA Emulation
Device 1.2
[opencl:cpu:1] Intel(R) OpenCL, AMD Ryzen 7 5800X 8-Core Processor  3.0
[2023.16.6.0.22_223734]
[ext_oneapi_cuda:gpu:0] NVIDIA CUDA BACKEND, NVIDIA GeForce RTX 3070 8.8 [CUDA 12.0]
```
(a)
```
$ acpp-info
================Backend information==================
Loadedbackend 0: HIP
Founddevice: Device 66a1
Founddevice: Device 66a1
Founddevice: Device 66a1
Founddevice: Device 66a1
Loadedbackend1: OpenMP
Founddevice: hipSYCL OpenMP host device

$ sycl-ls
================Backend information==================
[opencl:acc:0] Intel(R) FPGA Emulation Platform for OpenCL(TM), Intel(R) FPGA Emulation
Device 1.2
[opencl:cpu:1] Intel(R) OpenCL, Hygon C86 7185 32-core Processor 3.0 [2022.15.12.0.01_081451]
[ext_oneapi_hip:gpu:0] AMD HIP BACKEND, gfx906:sramecc-:xnack- 0.0 [HIP 40421.43]
[ext_oneapi_hip:gpu:1] AMD HIP BACKEND, gfx906:sramecc-:xnack- 0.0 [HIP 40421.43]
[ext_oneapi_hip:gpu:2] AMD HIP BACKEND, gfx906:sramecc-:xnack- 0.0 [HIP 40421.43]
[ext_oneapi_hip:gpu:3] AMD HIP BACKEND, gfx906:sramecc-:xnack- 0.0 [HIP 40421.43]
```
(b)

FIG. 1. Device discovery in the SYCL model: (a) is a local machine with an 8-core CPU and a NVIDIA GPU, (b) is a node of a supercomputer with a 32-core CPU and 4 AMD MI50.

*2. Queue initialization*

A queue represents a fundamental abstraction that orchestrates the execution of actions on a single device[57]. In SYCL, each queue is intrinsically linked to a specific device upon its creation, thereby simplifying the execution model and reducing the complexity of task management. The association of a queue with a single device at construction time allows for a clear and straightforward mapping of tasks to hardware resources, leading to more predictable performance characteristics. However, this also means that a queue's actions cannot be distributed across multiple devices, this inability to map a single queue to multiple devices can be a limitation in the era of heterogeneous computing. Future advancements in computing architectures may necessitate more dynamic and adaptable queue abstractions that can leverage the full potential of diverse and distributed computing resources.

### *3. Data management and transformation*

Accelerator devices often have their own attached memories that are not directly accessible from the host. Unified Shared Memory (USM)[57] plays a pivotal role in managing and migrating memories. USM simplifies the integration of accelerator devices with existing C++ codebases that utilize pointer-based memory management. It defines three types of memory allocations: *host*, *device*, and *shared*, with each satisfying specific requirements of data locality and accessibility. In the XFluids framework, USM is strategically employed to facilitate seamless memory operations. A minimal portion of data is designated as shared to prevent any potential degradation in performance that could arise from extensive shared memory usage. Instead, the flow-field state data is allocated in the device's local memory, which can be directly accessible by the accelerator device. This allocation strategy is optimized to exploit the high-speed memory available on the device, thereby enhancing computational efficiency and throughput. By allocating state variable data as device memory, XFluids ensures that the most frequently accessed data resides in the fastest accessible memory space, thus minimizing latency and maximizing performance.

### *4. Item assignment*

The ND-Range model is a sophisticated framework for explicit threads allocation in SYCL, similar to the thread hierarchy in CUDA and HIP. As depicted in FIG. **2**, the ND-Range is versatile, capable of being one-, two-, or three-dimensional (although only two-dimensional is shown here for simplicity), effectively mapping the work-items and work-groups within the iteration space. Each work-item within the ND-Range is analogous to a thread in CUDA and HIP and executes assigned tasks sequentially from inception to completion. This granular level of task assignment ensures that each work-item is fully utilized, corresponding to the thread execution in GPU computing models. In SYCL, work-groups are formed with a uniform number of work-items, which are typically optimized based on the target device's capabilities and the resource demands of each work-item. By aligning the work-group shape with the device's optimal execution parameters, SYCL's ND-Range model ensures efficient utilization of the device's resources.

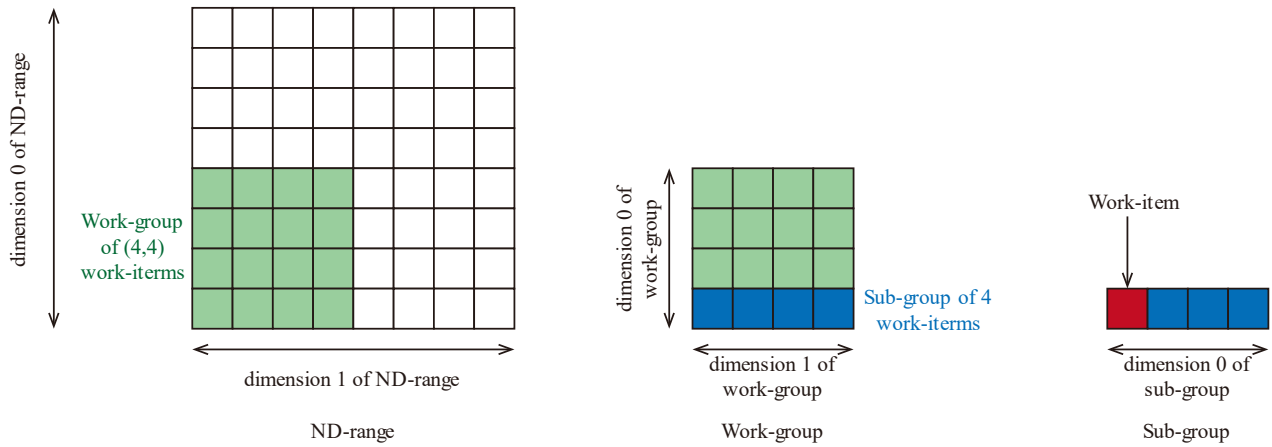

FIG. 2. Items assignment strategy in the SYCL model

### *5. Parallelism submission*

The *"parallel_for()"* function in SYCL, invoked by a handler object, *"q.submit([&](sycl::handler &h){h.parallel_for() {}};});"*, is used for assigning parallelism tasks to the execution queue, as illustrated by a demo code in FIG. **3**. This code is encapsulated within kernels, which are the fundamental building blocks for acceleration languages such as DPC++, SYCL, OpenCL, and CUDA. We observe that CUDA (or HIP) requires explicit annotations to identify device functions and kernels, including *"__global__"*, *"__device__"*, or *"__host__"*. These qualifiers serve as directives for the NVCC (or HIP) compiler to generate the appropriate code for the target GPU architecture. Unlike CUDA, one feature of SYCL is that it does not require specific compiler target identifiers to define and declare the kernels and device functions. Instead, the integration of Lambda expressions in SYCL provides a straightforward way to handle this, which simplifies parallelism submission, and integrates closely with modern C++ standards and idioms. With this, SYCL abstracts away the complexities associated with device-specific programming, offering a higher-level approach to parallelism that can automatically adapt to various hardware architectures.

```cpp
// Host serial code
for (int i = NumHaloX - 1; i < NumInnerX + NumHaloX; i++)
    for (int j = NumHaloY; j < NumInnerY + NumHaloY; j++)
        for (int k = NumHaloZ; k < NumInnerZ + NumHaloZ; k++){
            ReconstructFluxX(i, j, k, h_fdata);
        }
}
// CUDA code
ReconstructFluxX<<<gridsize, blocksize>>>(d_fdata);
// HIP code
hipLaunchKernelGGL(ReconstructFluxX, gridsize, blocksize, d_fdata);
// SYCL code
q.submit([&](sycl::handler &h){
    h.parallel_for(sycl::nd_range<3>(global_ndrange_x, local_ndrange), [=](sycl::nd_item<3> index){
    int i = index.get_global_id(0) + bl.Bwidth_X - 1;
    int j = index.get_global_id(1) + bl.Bwidth_Y;
    int k = index.get_global_id(2) + bl.Bwidth_Z;
    ReconstructFluxX(i, j, k, d_fdata);
}); });

// CUDA/HIP Kernel code
__global__ void ReconstructFluxX(Real* fdata){
    // operation executed on CUDA/AMD GPUs
    RoeAverage_x(fdata);
}
// SYCL Kernel code
void ReconstructFluxX(int i, int j, int k, Real* fdata){
    // operation executed on HOST/SYCL Devices
    RoeAverage_x(fdata);
}

// CUDA/HIP Device code
inline __device__ void RoeAverage_x(Real* fdata){}
// SYCL/Host Device code
void RoeAverage_x(Real* fdata){}
```

FIG. 3. The programming style differences between SYCL and other models.

In summary, the parallelism model of SYCL, facilitated by the *"parallel_for()"* function and its support for Lambda expressions, represents a more accessible and flexible approach. CUDA's model, while more intricate, offers precise control over GPU resources, which can lead to optimized performance for applications that are tailored to NVIDIA's ecosystem. Both models have their merits and are suited to different scenarios, depending on the developer's objectives and the hardware landscape of the deployment environment.

```cpp
// oneAPI Reducer definition
auto o_plus = sycl::reduction(&(argus), sycl::plus<>());
auto o_maxm = sycl::reduction(&(argus), sycl::maximum<>());
auto o_minm = sycl::reduction(&(argus), sycl::minimum<>());
// AdaptiveCpp Reducer definition
auto o_plus = sycl::reduction(&(argus), sycl::plus<double>());
auto o_maxm = sycl::reduction(&(argus), sycl::maximum<double>());
auto o_minm = sycl::reduction(&(argus), sycl::minimum<double>());

// parallel reduction submission
q.submit([&](sycl::handler &h) {
   h.parallel_for(sycl::nd_range<1>(g_rg, l_rg), o_plus, o_maxm, o_minm, [=](nd_item<1> index,
   auto &t_plus, auto &t_max, auto &t_min) { //
       int id = index.get_global_id(0);
       t_plus += redu_space[id];       // plus reduction
       t_max.combine(redu_space[id]);  // maximum reduction
       t_min.combine(redu_space[id]);  // minimum reduction
   });}).wait();
```

FIG. 4. A parallel reduction sample code of SYCL.

*6. Parallel reduction*

A parallel reduction sample of SYCL's abstraction is demonstrated in FIG. **4** for oneAPI and AdaptiveCpp. Compared to CUDA or HIP, SYCL significantly simplifies the expression of reduction kernels, enabling developers to focus on their applications' logic rather than the underlying hardware's intricacies. The abstraction can dynamically select the most efficient reduction algorithm by considering the specific characteristics of the device, the data type, and the employed reduction operation. This adaptability offers a substantial advantage in heterogeneous computing environments, where devices may exhibit diverse capabilities and performance. Moreover, SYCL's approach to reduction semantics provides a high-level interface that leverages the unique features of each device, such as specialized hardware units or optimized memory hierarchies, to execute more effectively.

*7. Synchronous and asynchronous execution*

After throwing the parallelism submission to the queue, a SYCL handler object will determine the range of work items and their dependencies. The synchronization between the host with the device's execution of tasks is typically achieved through commands like *"q.wait()"*, *"q.submit().wait()"* , *"q.memcpy().wait()"*, which are an essential component that guarantees data integrity and coherence. This is particularly important in complex applications where subsequent computations may depend on the outcomes of previous processing. Moreover, this synchronization point is a testament to the controlled concurrency offered by SYCL. It allows developers to harness the power of parallel computing while maintaining a clear and orderly execution flow. The ability to specify dependencies, *"q.submit([&](sycl::handler &h){ h.depends_on();h.parallel_for(){}};});"*, further enhances this control, enabling fine-grained management of task execution order and resource utilization. In addition, SYCL is automatically asynchronous by neglecting any wait operations, which is much simpler than CUDA and HIP where a complex stream configuration is usually applied to achieve asynchronous computations.

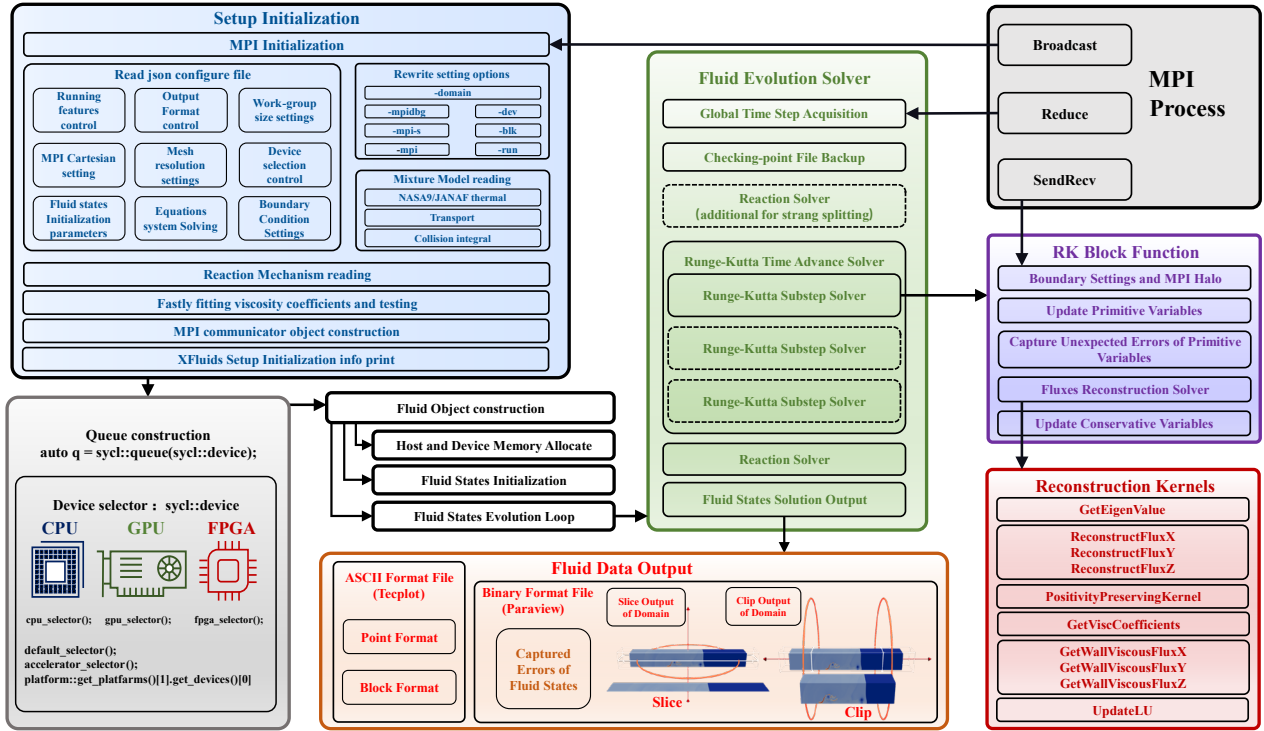

FIG. 5. The structure and solving process of XFluids.

## B. Solver structure and features

The operation workflow of the XFluids solver is illustrated in FIG. **5**, where the computational process is composed of four stages: initialization, solving process, data output, and memory management. FIG. **6** presents the timeline of a single computational step, while FIG. **7** illustrates the distribution of the time cost for each component of the solver. These three figures collectively reveal the sequence of function execution within the XFluids, and the time fraction of each part within the overall computation.

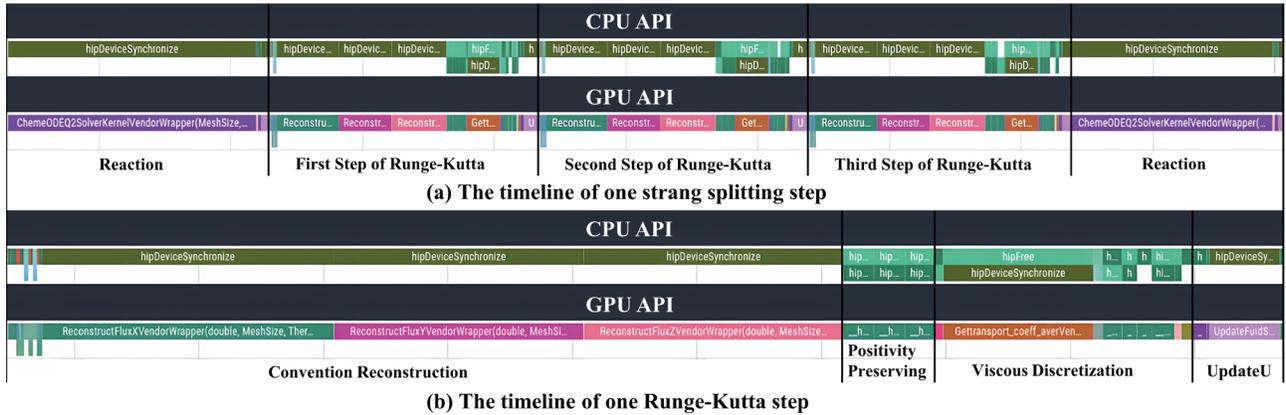

FIG. 6. XFluids HIP timeline in the case of Strang splitting: (a) one reaction step; (b) one Runge-Kutta step.

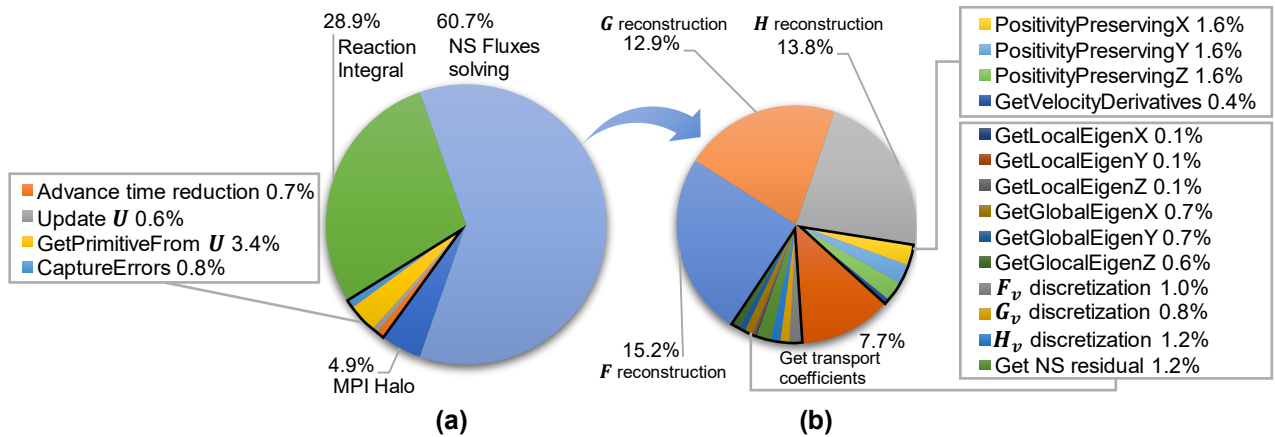

FIG. 7. XFluids time distribution map consists of 7 parts is shown in subfigure (a), including MPI halo treatment, reaction integral, time-step reduction, conservative variables $U$ update, primitive variables updated from $U$, unexpected errors catching, and NS fluxes calculation; Subfigure (b) details the time distribution of NS fluxes calculation, including local eigenvalue computation, global eigenvalue computation, the convention flux $F$, $G$, $H$ reconstruction, positivity preserving treatment for the convection fluxes, the velocity derivatives approximation, viscous transport coefficients calculation and discretization of viscous flux $F_v$, $G_v$, $H_v$, updating the NS residual of conservative variables.

### *1. Compile-time options*

Certain options of XFluids are available at the compile-time level, necessitating recompilation upon any modification to these features. This design choice is influenced by the compilation environment and the overarching goal of optimizing program performance. TABLE I lists some of the pivotal compile-time features alongside their respective functionalities. XFluids harnesses a Cmake-based framework for source code management and compilation, thereby enabling the manipulation of these features via Cmake options within the CmakeLists.txt file.

TABLE I. The Cmake compile options in XFluids.

| Cmake options | acceptable values | description |
| --- | --- | --- |
| SYCL_COMPILE_SYSTEM | oneAPI/OpenSYCL | using oneAPI or AdaptiveCpp compile system |
| SelectDv | host/cuda/hip | select the targeted platforms |
| ARCH | 75/80/906/1030, etc | the Compute Capability of targeting device |
| USE_MPI | ON/OFF | MPI support is enabled or disabled |
| AWARE_MPI | ON/OFF | GPU-enabled MPI support in XFluids is enabled or disabled |
| VENDOR_SUBMIT | ON/OFF | use CUDA/HIP native parallelism model or SYCL model |
| ARTIFICIAL_VISC_TYPE | ROE/LLF/GLF | artificial viscosity type |
| Visc | ON/OFF | enable the viscid flux terms of governing equations or not |
| Visc_Heat | ON/OFF | enable the Furrier heat transfer terms of viscid flux or not |
| Visc_Diffu | ON/OFF | enable the mass diffusion terms of viscid flux or not |
| VISCOSITY_ORDER | Second/Fourth | the numerical scheme order of the viscid flux terms |
| THERMAL | NASA/JANAF | fit gas thermodynamics using NASA or JANAF polynomials |
| COP | ON/OFF | solve the multi-component or single-component NS equations |
| ESTIM_NAN | ON/OFF | catch the unexpected errors or not during simulations |
| ERROR_OUT | ON/OFF | output intermediate variables upon errors captured or not |

Within these options in TABLE I, the *"SYCL_COMPILE_SYSTEM"*, *"SelectDv"*, and *"ARCH"* options form the foundation of the compilation system for XFluids, steering it towards the suitable platform and backend that aligns with the desired computational capabilities. The MPI and MPI-enabled features are controlled by *"USE_MPI"* and *"AWARE_MPI"*. The next option, *"VENDOR_SUBMIT"*, is used under the AdaptiveCpp SYCL compilation system to utilize the native parallel kernel functions and task submission models on NVIDIA or AMD devices. The Cmake options here also include the numerical method-related compile-time features. For instance, the *"ARTIFICIAL_VISC_TYPE"* setting determines the artificial viscosity in the convection flux reconstruction. In addition, the shear stress, Fourier heat transfer, and mass diffusion of the viscid flux terms are controlled by *"Visc"*, *"Visc_Heat"*, and *"Visc_Diffu"*, respectively, with their approximation accuracy being adjustable through the *"VISCOSITY_ORDER"*. To achieve high precision approximation and extensive temperature range of gas thermodynamic properties, the *"THERMAL"* setting is conventionally configured to NASA's thermodynamic polynomials by default, although the JANAF polynomials are also supported. To ensure the robustness of program execution, *"ESTIM_NAN"* and *"ERROR_OUT"* are typically enabled, facilitating the detection of potential errors and the generation of corresponding output files. Although this may result in increased memory consumption, the impact on program performance is marginal, as evidenced by FIG. 7.

TABLE II. The runtime features in XFluids.

| Runtime options | value type | sample | functionality |
|---|---|---|---|
| -domain | float | -domain=1.0,1.0,1.0 | the value lists computational domain size: length, width, height |
| -run | int | -run=400,400,400,10 | the value lists difference resolution and evolution steps: x_resolution, y_resolution, z_resolution, steps |
| -blk | int | -blk=8,4,2 | the value lists work-group shape for parallelism submission |
| -mpi | int | -mpi=2,1,1 | the value lists the MPI cartesian topology shape |
| -mpidbg | int | -mpidbg=1 | the value is given to open MPI attach debug mode |
| -dev | int | -dev=4,1,0 | the value lists the number and identity of the selected device |

### *2. Runtime features*

To decrease the possibility of recompilation for different flow simulations, XFluids' runtime features can be set by both the configuration file and common lines. The configuration file, typically in JSON format, currently contains 9 parameters as shown in the upper left corner of FIG. 5. Specifically, *"Output Format Control"* establishes data protocols and granularity catering for various post-processing software, e.g. Paraview, Tecplot. *"Work-Group Size Settings"* determines the block size and the specific thread shape for parallel task submission. *"MPI Cartesian Setting"* corresponds to the communication patterns among different mesh blocks, while *"Mesh Resolution Settings"* specifies the geometry and grid points of the computational domain. *"Fluid States Initialization Parameters"* configures the initial conditions of the flow field whose boundary conditions are prescribed in *"Boundary Condition Settings"*. *"Equations System Solving"* determines the type of the governing equations, including the existence of source terms, the application of the positivity-preserving method, and other numerical methods. Other parameters such as the maximum number of evolution steps and the Courant-Friedrichs-Lewy (CFL) number, etc. are included in *"Running Features Control"*, etc.

In addition to the configuration file, XFluids leverages options settings to override default configurations with command-line input. The accepted options are listed in TABLE **II**, which include the following. *"-domain"* adjusts the size of the physical computational domain. *"-run"* modifies grid resolution and evolution steps, tailoring the simulation to specific requirements. *"-blk"* is used to alter the work-group shape, while *"-dev"* redirects the program to target backend devices. *"-mpi"* reshapes the MPI cartesian topology, and *"-mpidbg"* activates a debug mode for the MPI version.

### *3. Setup initialization*

In XFluids' computational framework, the queue generation and device selection are critical steps that follow the initialization of parameters by the *"Setup"* submodule shown in the bottom left corner of FIG. **5**, which parses the configuration in the JSON file. In a multi-processing mode, each MPI processor selects its device and generates a corresponding queue, with the master processor reading the configuration file and broadcasting it to all other processors. Additionally, the *"Setup"* submodule is dedicated to preprocessing the chemical reaction models and initializing the flow field. After that, the data is transferred to the devices to ensure that the solver has the necessary information for the time evolution. The thermodynamic properties of the components are determined within the *"Thermal"* struct, while the parameters of the chemical reaction models are contained within the *"React"* struct.

### *4. Fluid object*

The complete solving procedure is shown in the left part of FIG. **5**, which is executed by a *"Fluid"* object for each MPI processor. This object is responsible for managing host and device memory allocation and initializing the fluid flow field based on the parameters defined in the configuration file. The timing step size for the reaction ODE solver is determined by the selected Lie or Strang splitting method, along with the globally acquired advancing time. The fluid dynamics equations utilize a three-step SSP Runge-Kutta method[40] for time integral, with specific boundary conditions and MPI halo treatment. After updating the primitive variables with the conservative ones, the fluxes reconstruction solver (as shown green part in FIG. **5**) is employed to reconstruct the inviscid cell-face fluxes using high-accuracy numerical schemes. This is followed by the application of positivity-preserving method and the discretization of viscosity fluxes. The conservative variables are then updated for the Runge-Kutta step. Upon completion of the three-step Runge-Kutta method, the solver executes the reaction integral and may require additional time marching of the fluids if the Strang splitting method is selected.

#### 5. Data output format

As shown in the bottom center of FIG. **5**, In XFluids, the data output is managed by the struct "OutFmt". To save the data storage space for large-scale simulations, this struct supports outputting both the entire computational domain and the manually defined partial domain, using either ASCII or binary file format. For instance, the string value *"0.000020: {-C=X, Y, 0.0; -P=yi[Xe] > 0.01, rho > 1.0; -V=rho, P, T, vorticity, yi[Xe]}"* within the JSON configuration file controls a sample of the output. The options *"-C"*, *"-P"*, and *"-V"* appended to the output physical time *"0.000020"* offer additional control, i.e. *"-C"* specifies the output of a slice of the entire computational domain, and is normal to the Z-axis direction, with the coordinate of z=0. *"-P"* manages the output of a clip of the entire computational domain determined by one or more criteria, which are used or logically connected within the program. *"-V"* lists of variables to be included in the output file. Other computational variables, although still reside in the memory, will be ignored during the output process. In such a way, the output management module ensures that only relevant data is extracted and stored, which optimizes the use of storage resources in the supercomputing centers.

#### 6. Checkpoint-based and error-catching computation

XFluids is designed to facilitate checkpoint-based computation, which may save a checkpoint file at the onset of each evolution step according to specific rules. When the program initiates, if it detects these files in the output directory, it will resume computation from the saved state of the flow field within those files. To optimize storage cost, the checkpoint only contains necessary conservative variables. By incorporating this checkpointing mechanism, XFluids enhances its robustness and allows efficient resource management and error handling, which is particularly beneficial in large-scale long-time simulations where unexpected interruptions usually happen. Moreover, XFluids continually monitors unexpected computational values, such as non-numeric values, infinities, and negative primitive variables like density, mass fraction, pressure, and temperature. Once any of such errors are detected, *"MPI_Bcast()"* is utilized by the error-catching processes to broadcast a termination signal to all other processors, preventing MPI runtime errors caused by the discrepancy of some processors encounter errors and exit prematurely with other processors continuing to run, ensuring all processors conclude their execution harmoniously. Additionally, XFluids outputs the intermediate variables that may be related to generating the errors, ensuring any potential issues are traced and analyzed. The execution locations of these two modules within XFluids can be found in the "Fluid Evolution Solver" part and the "RK Block Function" part of FIG. **5**.

### C. MPI implementation

Within the XFluids framework, MPI[58] is used for distributing computational tasks across multiple processors that are responsible for handling complex and resource-intensive calculations. By leveraging MPI, XFluids facilitates efficient data communication between processors, enabling them to work collaboratively and effectively.

### 1. Mesh topology

The organization of MPI processors is governed by a cartesian topology in the XFluids framework, which facilitates the structuring of processors into one-dimensional (1D), two-dimensional (2D), or three-dimensional (3D) grids. The distribution of MPI processors across the cartesian axes: $x$, $y$, and $z$, is defined by the external input parameters $mx$, $my$, and $mz$. Consequently, the total number of MPI processors, denoted as $N$ is the product of these parameters, $N = mx \cdot my \cdot mz$. Each MPI processor within this topology is uniquely identified by a rank number, which ranges from 0 to $N-1$, and is assigned a specific coordinate that denotes its position within the Cartesian grid. The initialization work of the fluid states varies for each processor based on its cartesian coordinate and the boundary conditions imposed on the solution domain. By leveraging this topology, XFluids ensures a coherent and efficient parallel processing environment, optimizing both the initialization phase and the ongoing computational operations within the simulation.

### 2. MPI reduction

The function *"MPI_Reduce()"* is used to effectively aggregate the local time steps computed by each subdomain, and to execute a reduction operation to determine the minimum global time step. This step is required to ensure the consistent advancement of the simulation for the entire flow domain. By obtaining the minimum time step, all processors proceed in lockstep, adhering to the CFL condition, which is essential for the numerical stability and accuracy of the simulation. Additionally, *"MPI_Reduce()"* is employed to calculate the maximum eigenvalue necessary for the global Lax-Friedrichs artificial viscosity.

### 3. Boundary halo

Boundary halo is used to exchange data between neighboring processors to ensure the consistency of the calculation. FIG. **9** illustrates an implementation of boundary halo supervised by the MPI process through transfer and receive buffers while two MPI processors are allocated in a 1D MPI Cartesian topology on a 2D solution domain. Each subdomain's flow field is divided into three regions: inner, border, and boundary. The inner region contains the data that are computed by the local MPI processor. The border region contains the data that are exchanged with the neighboring MPI processors through *"MPI_Sendrecv()"* operations. The boundary region contains the data that are received from the neighboring MPI processors or set by the physical boundary conditions.

### 4. GPU-enabled capability

The GPU-enabled MPI[59], such as OpenMPI integrated with NVIDIA's HPC SDK, and OpenMPI 5.0 with UCX support, improves the data transfer between different devices. These advanced MPI versions enable a direct communication pathway between device memory buffers, effectively eliminating the need for host memory buffers, as shown in FIG. **9**. This direct buffer communication is further enhanced by the utilization of high-bandwidth hardware interconnectors. This functionality is seamlessly supported by XFluids, which accommodates both traditional and GPU-aware message passing methods, thereby offering a versatile way for conducting frequent data transfer during large-scale simulations. Bypassing the traditional memory copying steps, a significant reduction in the message passing latency among MPI processors is proven to be achieved[60,61]. This streamlined approach not only reduces the time overhead but also minimizes potential bottlenecks in data transfer. As modern CFD programs grow in complexity and size, the ability to maintain high-speed communication without the constraints of host memory transfer becomes increasingly beneficial.

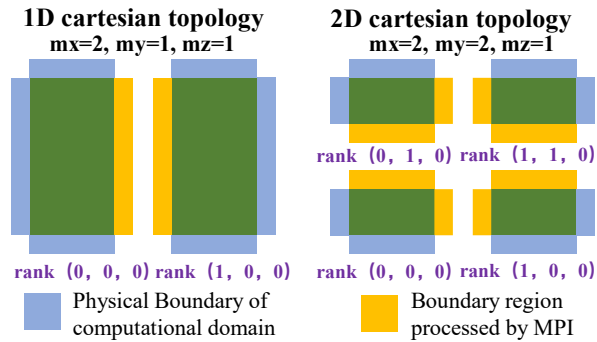

FIG. 8. A sample of 1D and 2D MPI cartesian topology.

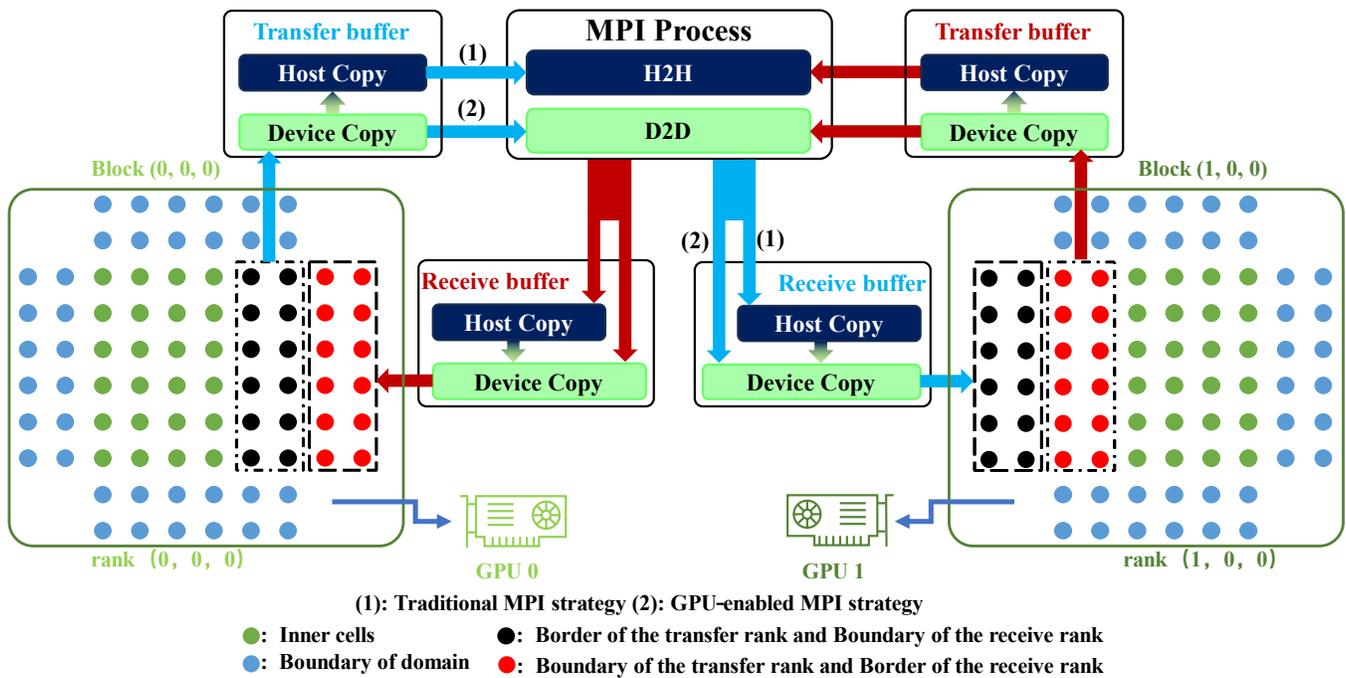

FIG. 9. Boundary halo treatment of two interconnected ranks in a 2D domain.

## D. Optimization strategy

First, flow-control idioms (*"if"*, *"switch"*, *"do"*, *"for"*, *"while"*, etc.) within kernel functions may lead to branch divergence among work-items in a sub-group, negatively impacting instruction throughput. To alleviate this issue, XFluids strategically replaces flow-control idioms with arithmetic operations (such as *"ceil()"*, *"floor()"*, and *"step()"*) wherever feasible.

Furthermore, for the kernel functions (including solving eigensystems, reconstructing convection fluxes, discretizing viscosity fluxes, and enforcing positivity-preserving functions) in different coordinate directions, the code snippets have been refactored as the unified formulations, or have been defined as macro functions, to improve the code readability and devolvement efficiency. Apart from these, the following parts list several optimization techniques related to the algorithm in XFluids.

TABLE III. XFluids GPU execute time with ARA optimization and without ARA optimization.

| Hotspot Kernels | work-group shape | | | | | | ARA |
| --- | --- | --- | --- | --- | --- | --- | --- |
| | (4,4,1) ratio[a] | (8,4,1) ratio | (8,8,4) ratio | (8,8,1) ratio | (4,4,4) ratio | (16,4,4) ratio | work-group shape |
| Updaterhoyi | 204.8% | 271.2% | 322.0% | 318.7% | 324.8% | 319.4% | (8,1,1) |
| EstimateYi | 145.4% | 111.3% | 178.8% | 153.4% | 194.4% | 185.5% | (8,1,1) |
| UpdateFuidStates | 104.4% | 124.8% | 218.9% | 159.8% | 163.7% | 221.2% | (8,2,1) |
| EstimatePrimitiveVar | 417.6% | 220.7% | 181.3% | 176.5% | 173.9% | 161.6% | (64,4,1) |
| ReactionODEKernel | 455.0% | 233.7% | 273.9% | 111.7% | 100.1% | 109.3% | (2,8,4) |
| GetLocalEigen | 118.1% | 111.6% | 108.5% | 103.9% | 115.8% | 108.2% | (32,1,8) |
| GetWallFluxX | 291.4% | 168.2% | 102.2% | 103.4% | 101.9% | 104.9% | (4,64,1) |
| GetWallFluxY | 249.2% | 151.4% | 103.8% | 100.6% | 102.6% | 103.5% | (16,4,1) |
| GetWallFluxZ | 252.4% | 154.4% | 104.9% | 111.3% | 100.9% | 104.7% | (8,2,8) |
| GetViscousCoeffs | 396.5% | 198.7% | 100.8% | 100.0% | 100.2% | 104.6% | (8,4,2) |
| UpdateFluidLU | 134.8% | 187.9% | 182.1% | 188.8% | 182.9% | 178.1% | (4,1,2) |
| EstimateLUKernel | 111.0% | 134.5% | 172.6% | 192.0% | 193.3% | 187.2% | (4,1,4) |
| UpdateURK3rd | 133.5% | 234.3% | 275.0% | 273.5% | 277.4% | 272.9% | (4,1,2) |

[a] ratio is a quotient calculated by dividing the running time of the specific work-group shape by the ARA optimization work-group shape running time.

### 1. Adaptive range assignment

For each kernel function, its work-group should be large enough to enhance overall multiprocessor occupancy. However, the size is also limited by the register usage of the algorithm inside the kernel function. In XFluids, we introduce the adaptive range assignment (ARA) treatment, i.e. it systematically evaluates all feasible work-group sizes for each kernel function and selects the most effective work-group shape to achieve optimal GPU performance. We conduct the ARA test on the AMD MI50 with $192^3$ 3D mesh to determine the optimal work-group sizes for the selected hotspot kernel functions. As shown in FIG. **10**, compared to the 6 different work-group configurations, the ARA's work-group shows the largest computational efficiency and dynamically adjusts its value for different hotspot kernel functions. Furthermore, in FIG. **10**, we compare the running timelines of the baseline and optimized ones through ARA. It is evident that, by utilizing the ARA, the reaction ODE solver function and the convection reconstruction function achieve a remarkable 46.4% and 39% performance improvement shown in FIG. **10**, respectively.

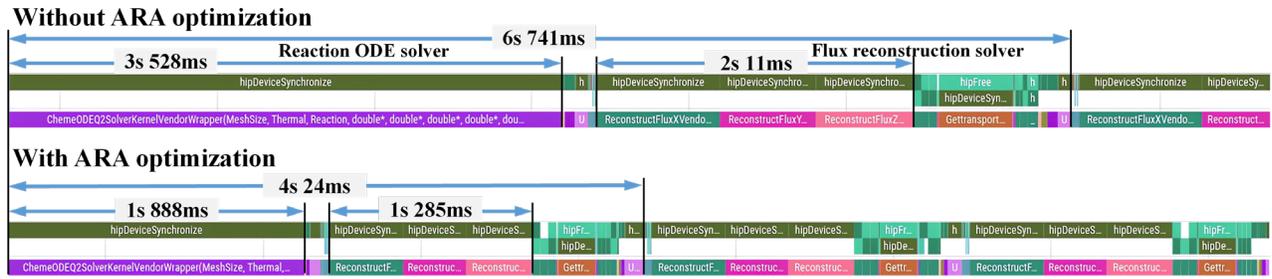

FIG. 10. Kernel functions timeline before and after ARA optimization.

*2. Hotspot device function optimization*

Numerous device functions contribute to the major computation complexity of the CFD solver. One of the most time costly ones is the nonlinear reconstruction scheme (e.g. WENO5) device function inside convection fluxes reconstruction kernels. To handle this computational bottleneck, we use 3 main optimization strategies as depicted in FIG. **11**. First, the frequently used coefficients are precomputed as device constants and stored in read-only constant memory. The use of constant memory significantly increases the access speed compared to the global memory, thereby mitigating memory latency. Second, the number of division operations, which are extremely slow on GPUs, is reduced by predefined operations and variables. By minimizing these costly operations, the overall efficiency of the GPU execution improves. Third, we also optimize the performance of the single-precision computations by avoiding the data format conversion. Directly running the two versions on NVIDIA GeForce 3070, we demonstrate the optimization effect of the WENO5 function and find a substantial 25.6% performance improvement.

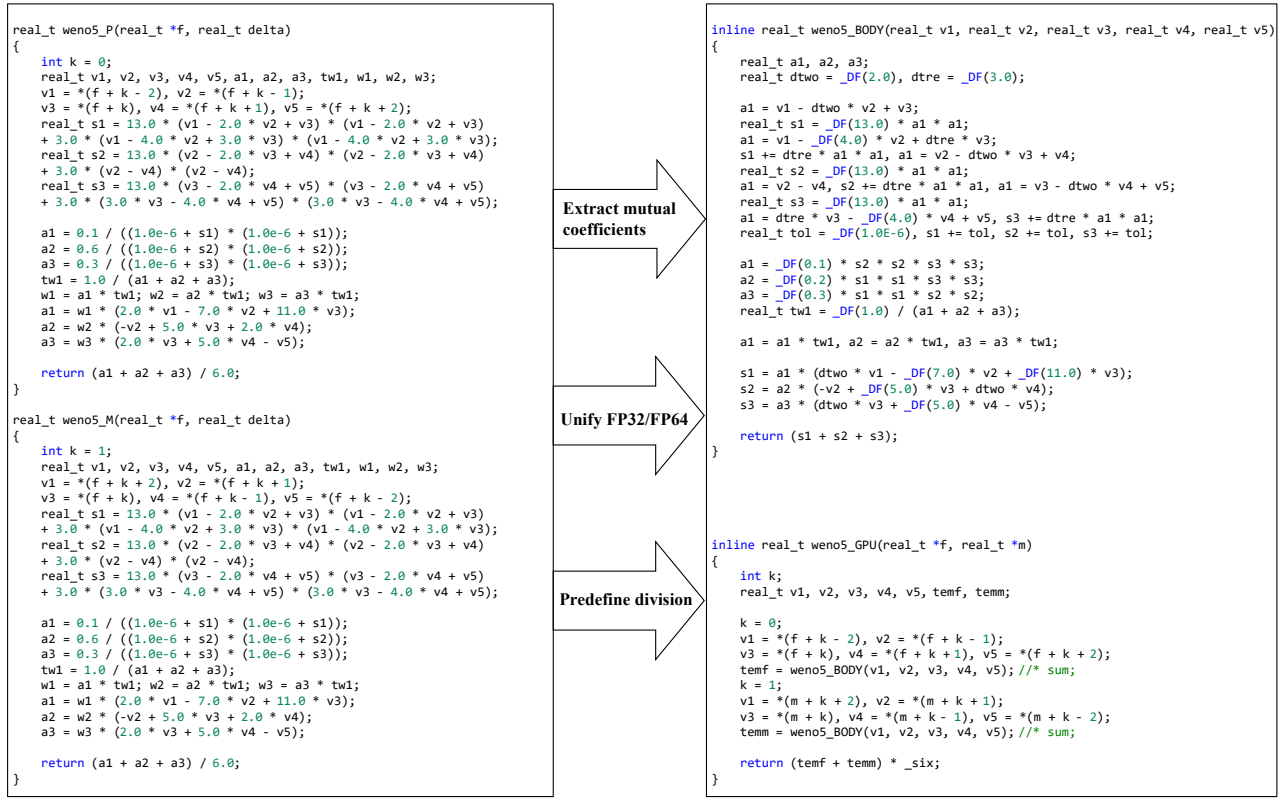

FIG. 11. GPU optimization of the hotspot WENO5 function.

### *3. Partial eigensystem reconstruction*

The eigensystem used for flux reconstruction at the cell faces consists of an eigenvalue vector and two eigenvector matrixes, see FIG. **12**. The traditional method solves the entire eigensystem at once (EEO) which requires a quadratic increase of the registers within the kernel function as the number of components increases. Consequently, as indicated in FIG. **13**(a), this significantly augments the register spilling, leading to the increase of local memory usage, which is significantly slow as it essentially is part of the global memory. Therefore, we modify the EEO by the partial eigensystem reconstruction (PER) method, i.e., for the n-th element of the flux vector, only the corresponding n-th eigenvalue, the n-th row of the left eigenvector matrix, and the n-th column of the right eigenvector matrix are computed at each step. Thus, the original eigenvector matrixes are replaced by two vectors, leading to a substantial reduction of the register usage, as illustrated in FIG. **13**(a). As a result, the PER method achieves significant acceleration over EEO during convection reconstruction, as shown in FIG. **13**(b).

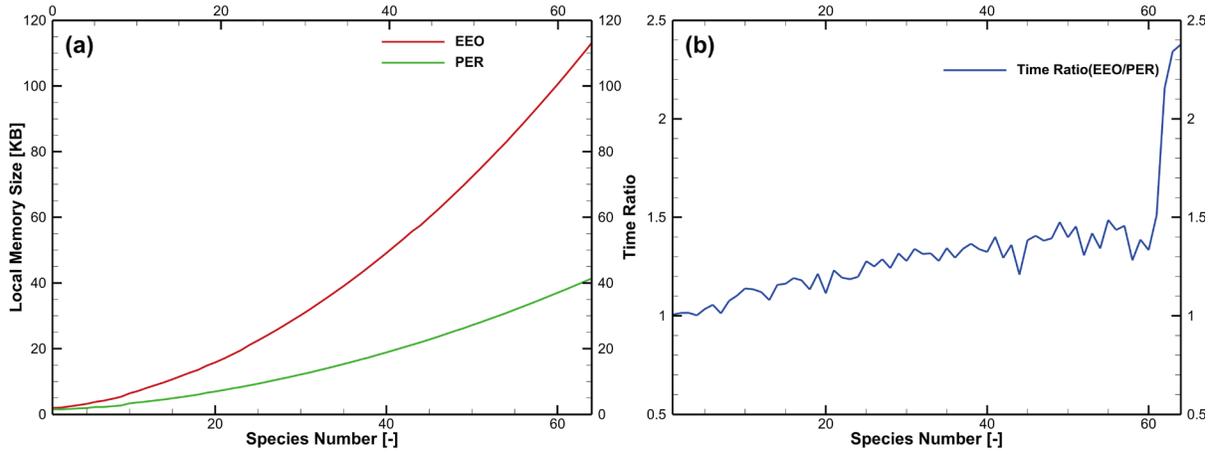

FIG. 12. The eigen decomposition used for multi-species flux reconstruction at the cell faces.

FIG. 13. (a) The spilled-out local memory size of two methods as species number increases. (b) The running time ratio of the full eigen system over the partial eigen matrix system as species number increases.

*4.* **Fastly fitted transport coefficients**

Direct calculating the gas transport coefficients according to the exact formulas in Section II.C.3 would consume excessive computing resources. Fortunately, as these coefficients are temperature-dependent functions, we can construct the following temperature-based polynomials[62] to fit these data,

$$\ln \eta_k = \sum_{n=1}^{N} a_{n,k} (\ln T)^{n-1}, \qquad (29)$$

Here $\eta_k$ represents $D_{st}$, $\mu_s$ or $\kappa_s$, and $a_{n,k}$ is the polynomials' coefficients. For each transport coefficient, e.g. the viscosity $\mu_s$, one first selects $N$ different temperatures and computes the $\eta_k$ by the exact formulas in Section II.C.3. After that, the coefficients of the polynomials can be solved by such a linear system and stored in the device memory before the computation of the fluid evolution. FIG. **14** demonstrates the high accuracy of this fitting formula by testing the transport coefficients of some common species $N_2$, $O_2$, $CH_4$, $CO_2$, and $OH$.

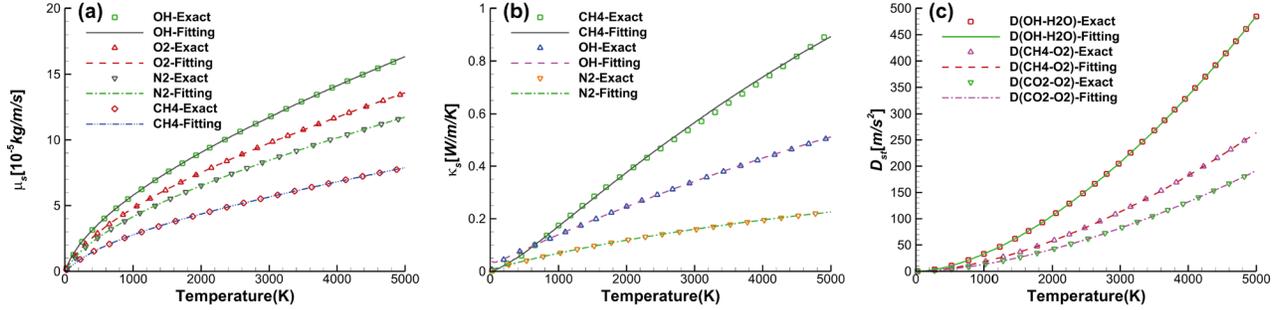

FIG. 14. Comparison of exact and fitted transport coefficients: (a) the viscosity, (b) the thermal conductivity, and (c) the binary diffusion coefficients.

## IV. NUMERICAL VALIDATION

### A. Multicomponent inert shock tube

In this case[33,63], the $0.1m$ long domain discretized with 400 grid points is evenly fulfilled with an inert multicomponent gas mixed by a molar ratio $X_{H_2} : X_{O_2} : X_{AR} = 2:1:7$ separated into left and right parts by an initial condition as $(T_L, P_L) = (400K, 8000Pa)$ and $(T_R, P_R) = (1200K, 80000Pa)$, with the left inflow and right outflow boundary condition imposed. We test the solver by using 3 different reconstruction schemes, WENO5, WENOCU6, and WENO7, and compare the distribution of the velocity, temperature, density, and ratio of specific heat capacity along the *x*-direction at 40 μs with the reference solution in Martínez et al.[53] As shown in FIG. **15**, although negligible discrepancy is observed near the discontinuities, a good consistency can be found for all the flow states. This demonstrates that the physical models and numerical methods of the convection portion are correctly implemented in the current solver.

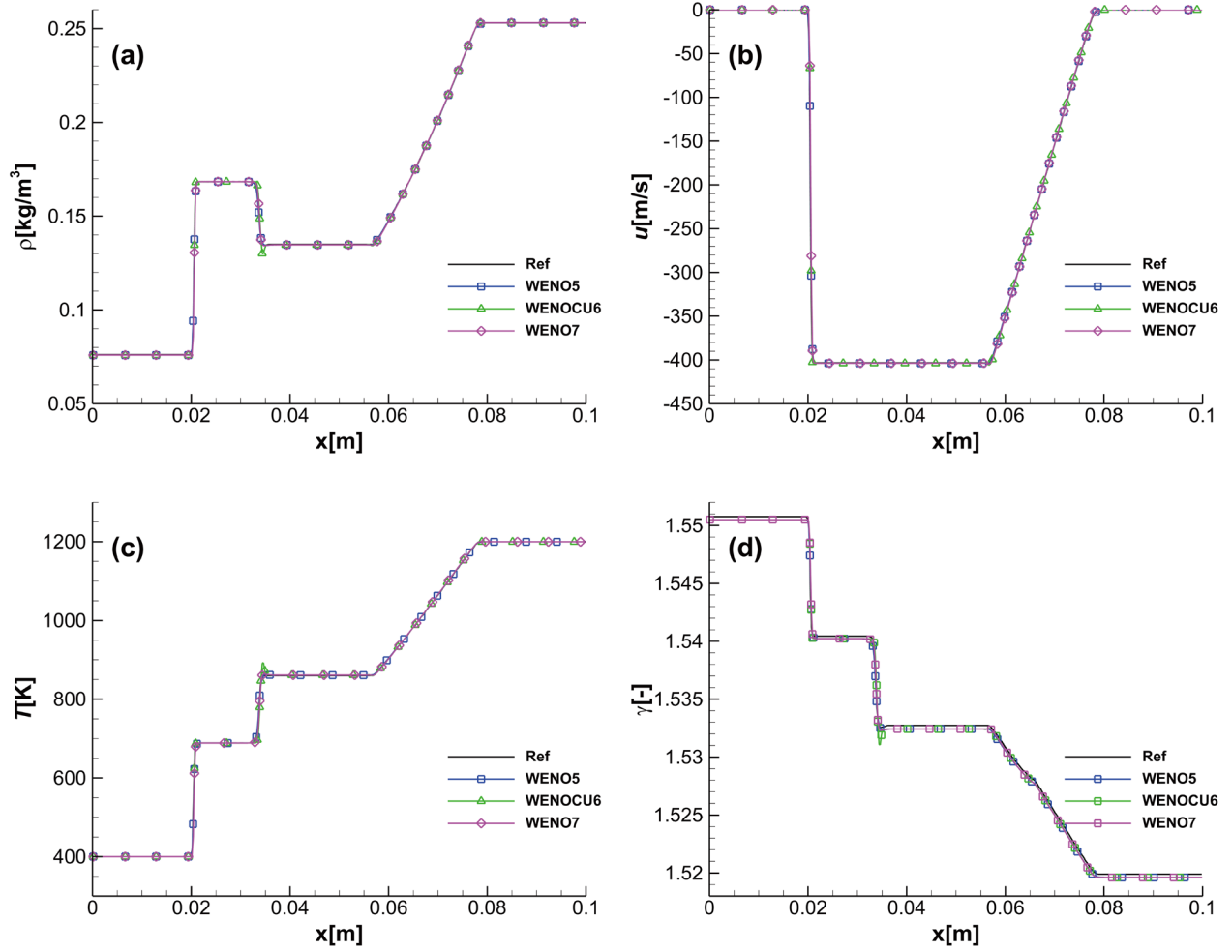

FIG. 15. (a) The density, (b) the velocity, (c) the temperature, and (d) the ratio of specific heat capacity along the tube at 40 μs.

**B. Multicomponent diffusion**

This case is performed to validate the molecular transport and heat conduction terms of XFluids. We solve a 1D multicomponent diffusion problem with a domain length of $l = 0.05m$, which is discretized using 200 uniform grid cells. The periodic boundary conditions are imposed on both sides of the domain. And the initial pressure and velocity are as $p = 1atm$, and $u = 0m/s$, receptively. The temperature and component mass fractions are given by

$$Y_k(x) = Y_{k_o} + \left(Y_{k_f} - Y_{k_o}\right) f(x), \quad k = O_2, N_2, H_2O, CH_4$$
$$T(x) = T_o + \left(T_f - T_o\right) f(x) \qquad , \qquad (30)$$
$$f(x) = 1 - 0.5 \exp\left(-(x - x_0)/d^2\right)$$

and expressions in TABLE **IV**. Here the parameters are set as $d = 0.0025m$ and $x_o = 0.025m$. The temporal evolution of the temperature and the mass fraction of $CH_4$ are depicted in FIG. **16** for different time steps using the WENO-CU6 reconstruction scheme. Clearly, the numerical results show a good agreement with the reference data in the literature[53], which demonstrates the accuracy of the heat conduction coefficient and mass diffusion coefficient calculations in our solver.

TABLE IV. The temperature and mass fractions on fuel and oxidizer sides.

| Fuel side | Oxidizer side |
|---|---|
| $T_f = 320K$ | $T_o = 1350K$ |
| $Y_{O_{2_f}} = 0.195$ | $Y_{O_{2_o}} = 0.142$ |
| $Y_{N_{2_f}} = 0.591$ | $Y_{N_{2_o}} = 0.758$ |
| $Y_{H_2O_f} = 0.0$ | $Y_{H_2O_o} = 0.1$ |
| $Y_{CH_{4_f}} = 0.214$ | $Y_{CH_{4_o}} = 0.0$ |

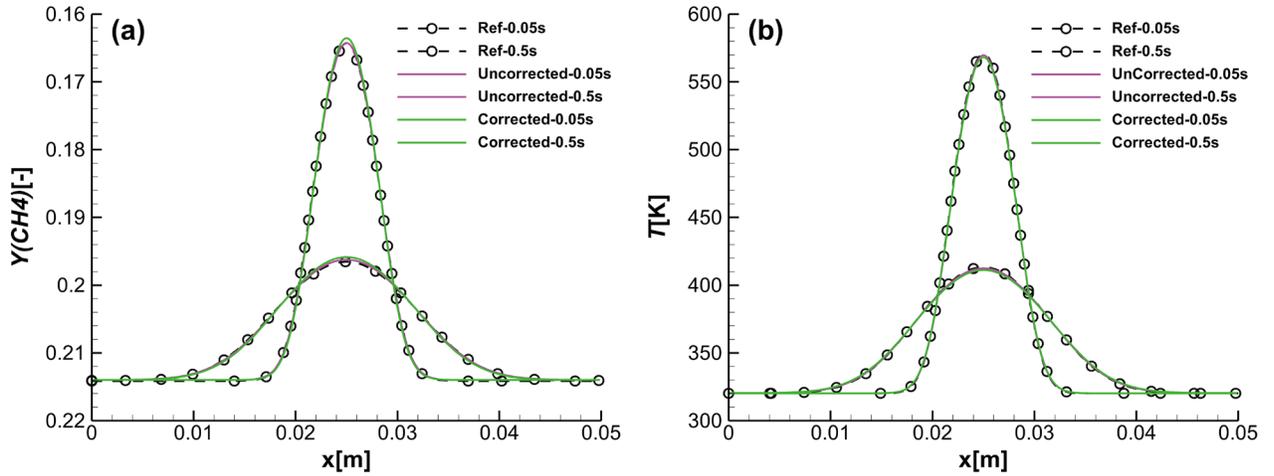

FIG. 16. (a) The mass fraction of $CH4$, and (b) the temperature distributions along the domain at 0.05s and 0.5s.

### C. Zero-dimensional nitrogen dissociation

The nitrogen dissociation is an exothermic phenomenon in which nitrogen molecules ($N_2$) break down into nitrogen atoms ($N$) under high temperature and pressure conditions. In this study, we consider a closed chamber that contains a mixture of $N_2$ and $N$ with an initial molar fraction ratio of $X(N_2):X(N) = 2:1$, a temperature of 4000 K and a pressure of $10^5$ Pa. This simplifies the problem to a purely chemical reaction progress, which serves as a benchmark to evaluate the accuracy of the reaction portion of XFluids. Assumed that the total mass and total internal energy of the mixture are conserved during the reaction process, we consider the $N_2$ dissociation reaction mechanism proposed by Park[64], which consists of two reversible reactions,

$$\begin{aligned} 2N_2 &\rightleftharpoons N_2 + 2N \\ N_2 + N &\rightleftharpoons 3N \end{aligned}. \tag{31}$$

The simulated results solved by XFluids and the anastomotic equilibrium mass fractions components predicted by CEA[65] are plotted in FIG. **17**, implying that consistency between XFluids and CEA is achieved.

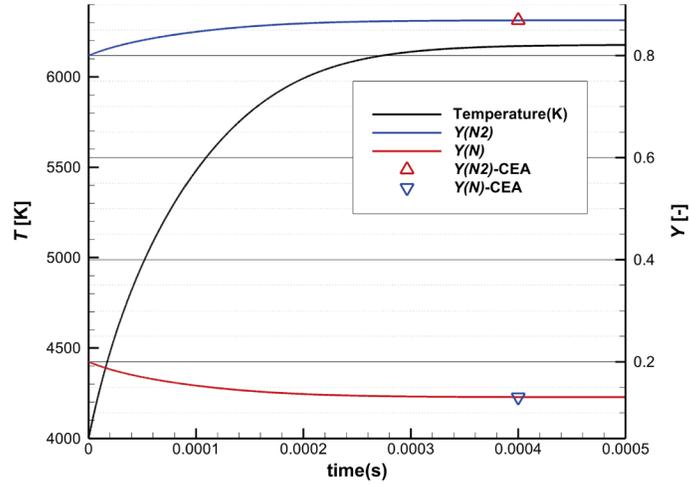

FIG. 17. Time evolution of the temperature and mass fraction of the reactants in the nitrogen dissociation problem.

**D. Reactive Shock Tube**

In this section, we use XFluids to present a numerical solution of the 1D reacting shock tube[53] problem to evaluate the accuracy of our high-order reconstruction schemes and chemical reacting source terms solver. The shock tube with a length of $0.12m$ is discretized by 400 structured mesh cells, initialized as a premixed mixture of hydrogen, oxygen, and helium with a molar ratio of $X_{H_2}:X_{O_2}:X_{AR}=2:1:7$. The initial flow state of the mixture on the left side and the right side are $(0.072 kg/m^3, 0 m/s, 7137 Pa)$ and $(0.18075 kg/m^3, -487.34 m/s, 35594 Pa)$, respectively. We impose a no-slip wall condition at the left boundary and an outflow condition at the right boundary. The 19 reactions $H_2$-$O_2$ mechanisms mentioned in Section II.C.4 are evaluated. As shown in FIG. **18**, the temperature, density, velocity, and hydrogen atom mass fraction at three time instants, 170 μs, 190 μs, and 230 μs, fit well with those obtained by Wang et al.[33], demonstrating the validation and accuracy of our approach.

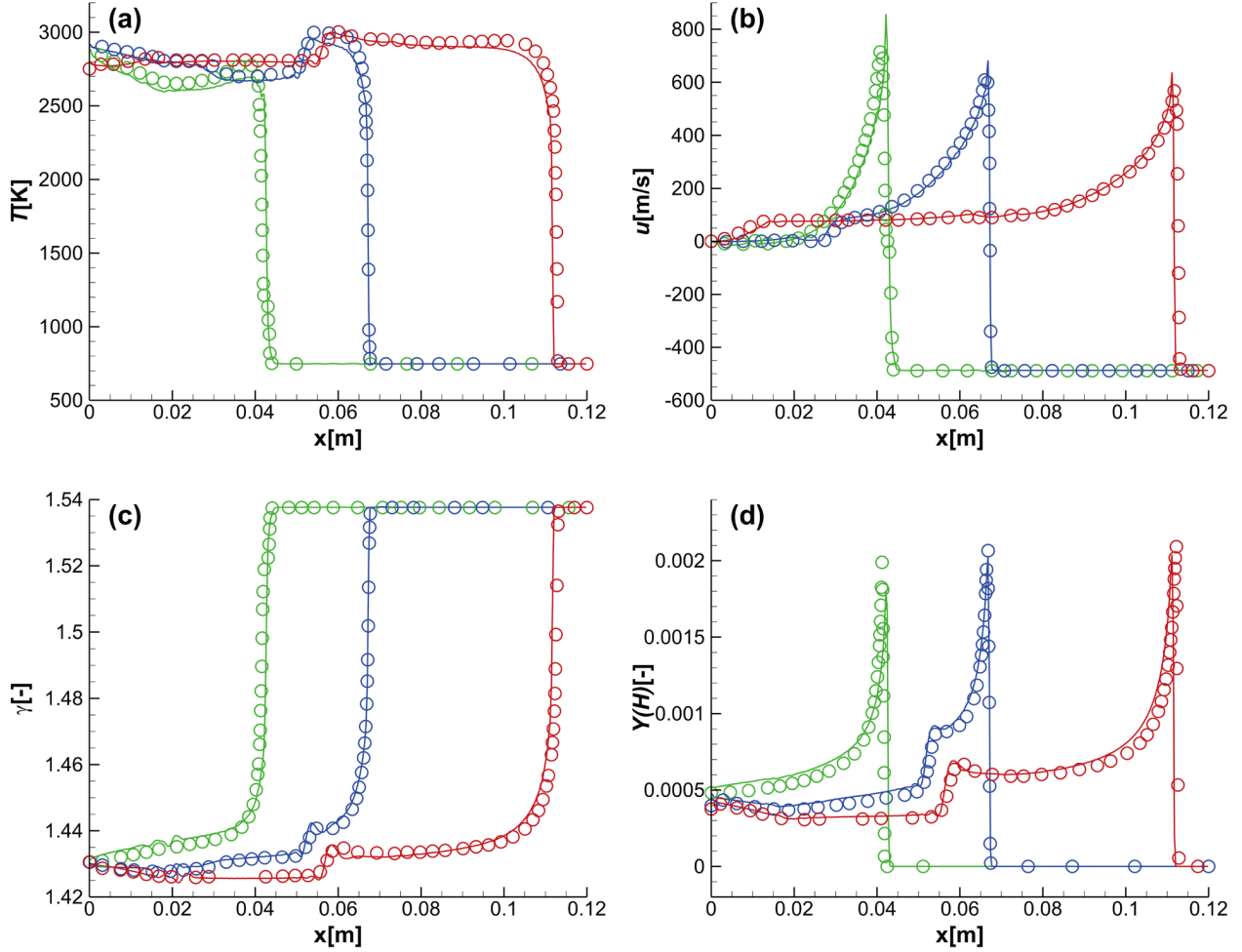

FIG. 18. Comparison of (a) the temperature, (b) the velocity, (c) the ratio of specific heat capacity, and (d) the hydrogen atom mass fraction of Wang et al.[33] (circle symbol) and XFluids (solid line) along the tube at 170 μs (green color), 190 μs (blue color), and 230 μs (red color).

## V. PERFORMANCE BENCHMARK

### A. Acceleration ratio of different backends

To demonstrate the cross-architecture feature of XFluids, we have selected 6 different GPUs and 5 CPUs from the mainstream vendors (Intel, AMD, NVIDIA), as listed in FIG. **19**. As mentioned previously, the SYCL programming model is compatible with other GPU programming models in terms of coding style. First, we measure the execution time of the 6 different GPUs and 5 CPUs by using the SYCL model with double-precision format for the 2D shock-tube problem. The acceleration ratio of each GPU card over one single CPU core is listed in FIG. **19**(a), which shows an acceleration ratio ranging from 29.5 to 453.3 for AMD and NVIDIA GPUs. This value can be relatively smaller when the desktop CPUs (AMD 5600X, AMD 7950X, etc.) are used as the baseline, as they usually have much higher single-core turbo frequency. In addition, the acceleration ratio is higher on data center GPUs which have higher double-precision performance. Note that the 2 Intel GPUs exhibit very low performance. This is because they lack double-precision hardware support, and we run XFluids on them by emulating the double-precision operations with Intel's OneAPI compiler, which is usually very slow. To demonstrate the high performance of XFluids, in FIG. **19**(b), we also compare the acceleration ratio (single GPU card over one single AMD 5600X CPU core) between the SYCL model in XFluids and the vendors' native model (CUDA and HIP). Compared to the existing programming models, SYCL significantly enhances the portability of XFluids, without deteriorating the performance.

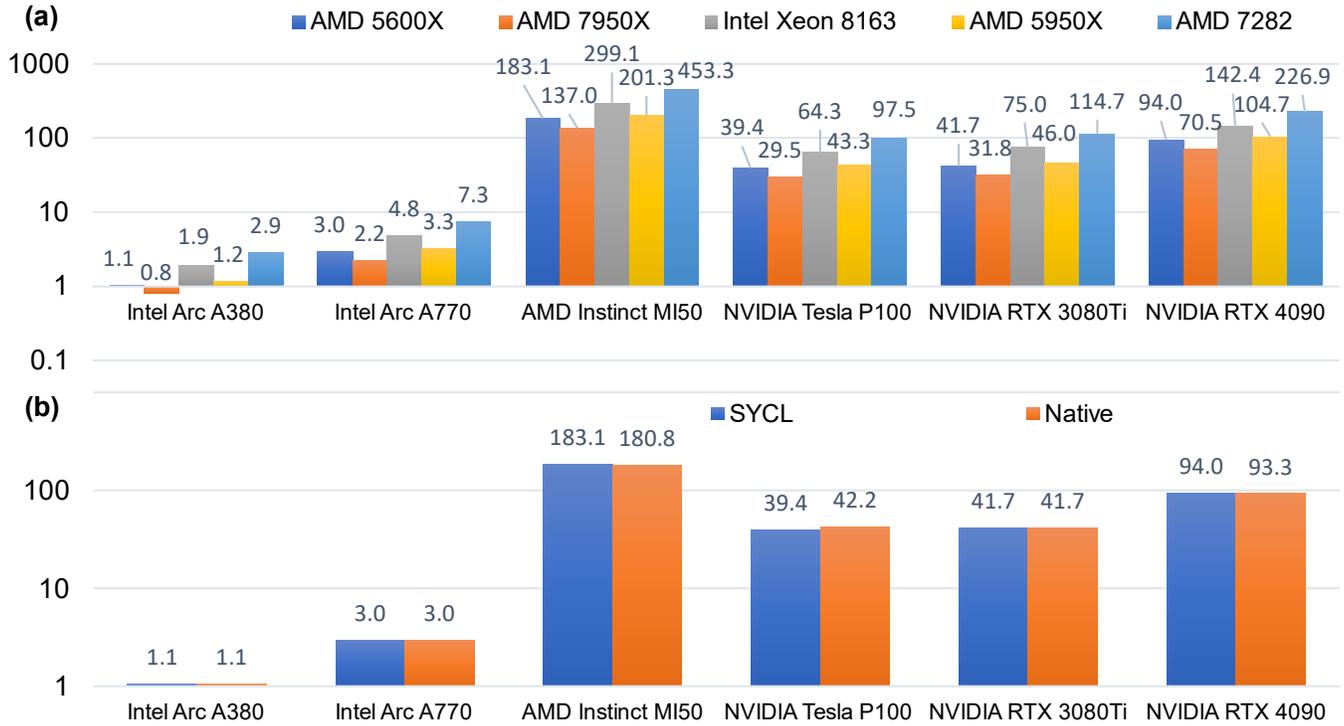

FIG. 19. (a) The acceleration ratio of different SYCL GPU backends to one single core of CPU backends and (b) the comparison of the acceleration ratio (versus one CPU core of AMD 5600X) between the native GPU programming models and the SYCL model.

**B. Parallel scalability**

After measuring the single-GPU acceleration ratio of XFluids, its multi-GPU performance is also tested in this section. As mentioned before, the MPI library is leveraged in XFluids to manage the parallel tasks and data transfer across a supercomputer. Here each compute node of this supercomputer is equipped with 4 AMD Instinct MI50 GPU and 32 CPU cores. In this case, the shock-bubble interaction (SBI) problem in the next section is used to compute the weak scalability of XFluids, which handles one GPU by one individual MPI processor. To maximize device memory utilization (each GPU has 16GB global memory), the 3D and 2D SBI cases use a grid resolution of $192^3$ cells and $3072 \times 2304$ cells, which consumes 15.66GB and 13.92GB memory, respectively. As depicted in FIG. **20**, the parallel efficiency of XFluids is considerably high, as its weak scaling exceeds 95% for both inert SBI (ISBI) and reactive SBI (RSBI) cases, with the finest resolution reaching the order of 7.2 billion cells. In addition, in FIG. **20**, the efficiency is higher for the reactive cases (orange and yellow lines) than the inert case (blue and gray lines), and higher for 3D cases (orange and blue lines) than 2D cases (yellow and gray lines) as well, as the additional computational resources introduced will the reduce proportion of communication time within the whole simulation time.

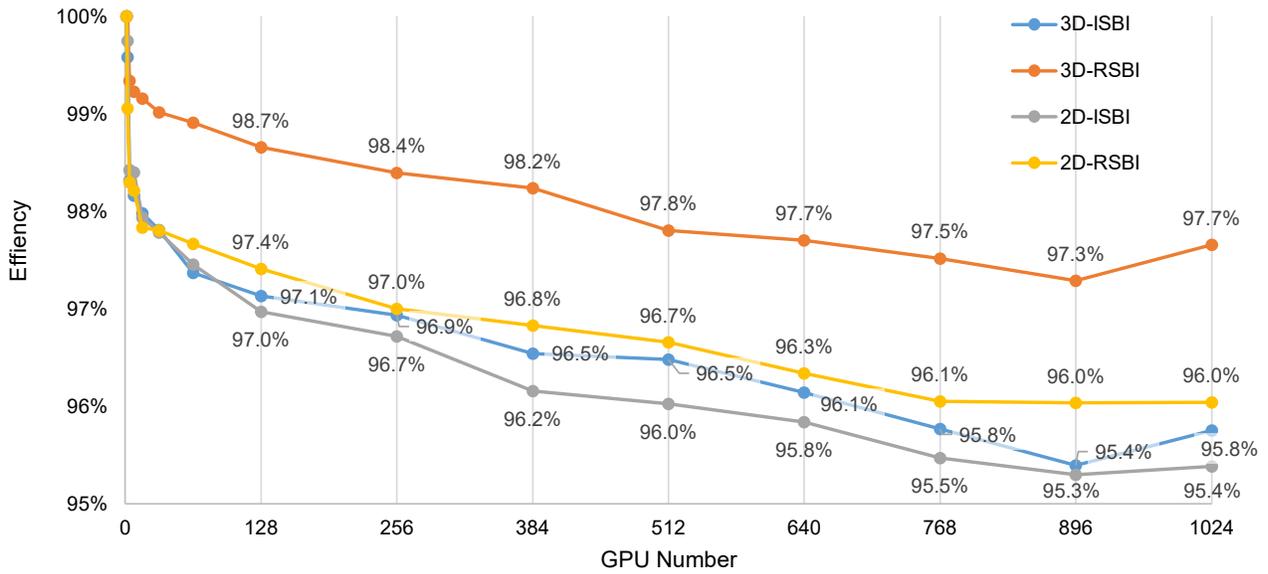

FIG. 20. The weak-scaling parallel efficiency of XFluids.

## VI. Applications to multi-component shock-bubble interactions

### A. Physical and numerical setups

Shock-bubble interaction (SBI)[1] is a fundamental problem that has attracted considerable numerical and experimental attention in the past several decades. In this case, the shock wave induces an initial compression and deformation of the bubble, which generates a pair of vortices in the 2D configuration or a ring of vortices in the 3D configuration. As the interaction evolves, depending on the strength of the shock wave and the density ratio between the bubble and the surrounding fluid, multiple secondary vortices may appear, and the bubble mass may be stripped from its original shape, followed by significant mixing between the bubble and the ambient fluids[1].

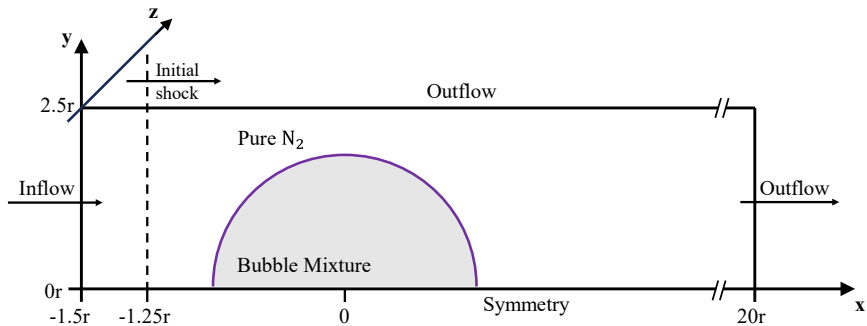

FIG. 21. The computational domain and boundary conditions of the SBI.

Based on the previous study[5,29,66,67], both inert SBI and reactive SBI problems in rectangular domains are simulated, with the domain size being (20$r$, 2.5$r$, 0.0$r$) for 2D cases and (20$r$, 2.5$r$, 2.5$r$) for 3D cases, where $r$ is the initial radius of the bubble.

As shown in FIG. **21**, this bubble contains a gas mixture and is surrounded by pure $N_2$. The interface between the bubble and $N_2$ is defined by an interface equation[67] in terms of the molar ratio of $N_2$ as

$$d = \sqrt{(x-x_0)^2 + (y-y_0)^2 + (z-z_0)^2} - r$$
$$X_{N_2} = \frac{(1-2\zeta)\tanh(\xi \cdot d) + 1}{2}, \quad (32)$$

where $(x_0, y_0, z_0) = (0,0,0)$ is the initial position of the bubble center, and $\xi$ is the sharpness coefficient. $\zeta$ is set at a tiny value for the robustness of the simulation, i.e. 0.0001 in our study, and leads to an initial non-zero value of the Molecular Mixing Fraction (MMF)[29,67]. The molar fraction ($X = 1 - X_{N_2}$) inside the bubble is distributed among the 3 gases, ensuring a stoichiometric mixture with molar fractions of $X_{O_2}:X_{Xe} = 51:49$ for the inert simulations and $X_{H_2}:X_{O_2}:X_{Xe} = 6:3:11$ the reacting simulations, keeping the Atwood number of the simulations as 0.49. Given the shock propagation Mach number $Ma = 2.83$, the shock wave is initialized to the left of the bubble at $x = -1.25r$. The pre-shock state is defined by $T_0 = 295K$, $P_0 = 1atm$, $\rho_0 = p_0 / (R_{N_2} T_0)$, $c_{N_2} = \sqrt{\gamma_{N_2} p_0 / \rho_0}$, and the initial post-shock state is obtained by the standard Rankine-Hugoniot conditions,

$$\rho' = \rho_0 \frac{(\gamma_{N_2}+1)Ma^2}{2+(\gamma_{N_2}-1)Ma^2},$$
$$u' = Ma \cdot c_{N_2}\left(1 - \frac{\rho_0}{\rho'}\right), \quad c_{N_2} = \sqrt{\gamma_{N_2} p_0 / \rho_0}, \quad (33)$$
$$p' = p_0\left(1 + 2\frac{\gamma_{N_2}}{\gamma_{N_2}+1}(Ma^2 - 1)\right).$$

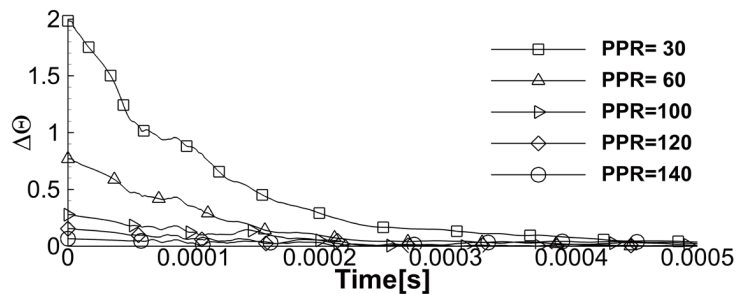

FIG. 22. The grid convergence study of MMF in RSI.

## B. Grid convergency study and numerical validation

Following previous research[67], the resolution of the simulation is quantified by the number of points per radius (*ppr*), and the sharpness coefficient is therefore represented as $\xi = 2l \cdot ppr$, with *l* being the domain length. As the error of MMF, $\Delta\Theta$, converges as *ppr* exceeds 140 in previous work[67], we select 6 resolutions, (30, 60, 120, 140, 160) *ppr* in this study. FIG. 22 shows the increase of *ppr* leads to a successive reduction of the error $\Delta\Theta$ which is defined as the difference between the simulated MMF of each *ppr* with the result of *ppr*=160. For the resolution of *ppr*=140, the error is reduced to less than 7% in the early stage, and less than 1% in the long-term evolution of MMF. This grid convergence study shows that a resolution of *ppr*=160 usually suffices for the 3D SBI problem, which is used in the following simulations.

The normalized transverse diameter (TBD), which is defined as $\widetilde{\Lambda}_y = \Lambda_y / D$, describes the maximum spatial expansion of the bubble in the transverse direction ($\Lambda_y$) normalized by its initial diameter $D$. The bubble diameter $\Lambda_y$ is measured based on a threshold value of the mass fraction of $Y_{Xe} = 0.01$. O'Conaire's $H_2$-$O_2$ combustion reaction mechanism[52] mentioned in Section II.B is applied in the reacting case, and its validation and modification for shock-bubble interaction have been shown in previous study[67]. The validation of our simulation is provided in FIG. 23 by comparing with the reference data, indicating a satisfactory agreement between our 3D simulation result with both the experiment data[5] and previous 3D numerical data[67].

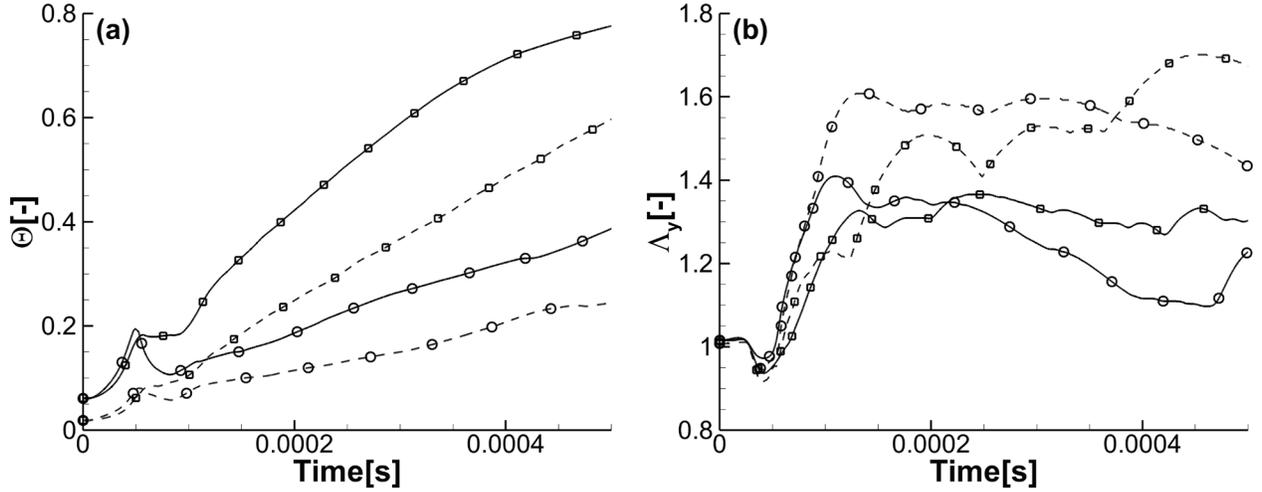

FIG. 23. The TBD of the bubble over time: (a) ISBI and (b) RSBI. Here the solid line presents our simulation by using a resolution of *ppr*=160 for 3D SBI and *ppr*=320 for 2D SBI. The symbols correspond to the ISBI experiment data of Haehn et al.[5] in (a) and the RSBI simulation data of Diegelmann et al.[67] in (b).

## C. The reacting and 3D effect on bubble deformation and mixing process

To investigate the reacting and 3D effects on the mixing and deformation process in the SBI problem, we have illustrated the temporal evolution of TBD and MMF for SBI in FIG. 24, and also time evolution of the temperature contours in FIG. 25 as well. Initially, the molecular mixing fraction of both 2D and 3D cases in FIG. 24(a) shows a sudden increase during the shock wave passage in the early stage (t < 50 μs). Subsequently, it remains approximately invariant for ISBI and decreases for RSBI during the shock focusing (50 μs < t < 100 μs), respectively. Then the mixing level rises again, where the 3D cases show a much higher slope than the 2D cases, corresponding to a higher overall mixing rate. The reaction wave that exists in the reacting cases is a supersonic detonation wave[1], which suppresses the growth of secondary instabilities and reduces the mixing by up to 50% compared to the inert simulation, which is in agreement with the previous study of MMF by Diegelmann et al.[67] In the long-time evolution, a notable transverse expansion is observed in 2D simulations of FIG. **24**(b) and FIG. **25**(a, b), which arises due to the absence of the vortex stretching effect[66]. In contrast, when extending to 3D simulations, the growth rate of the vortex stretching increases significantly and therefore contributes to the decrease of the transverse bubble diameter. Thus, the main difference between 2D and 3D results, which is governed by the interplay between the transverse expansion and the vortex stretching, is correctly reproduced in FIG. **24** and FIG. **25**.

Apart from the vortex stretching effect in 3D configurations, the reactive effect also significantly alters bubble defamation and mixing in SBI. As reported previously[67], in the RSBI, the passage of the shock induces strong compression of the bubble gas and a sudden change in the thermodynamic properties, which are sufficient to ignite the reactive gas mixture, as shown in the 2nd time snapshots of FIG. **25**(b). In this condition, the shock-focusing within the bubble is enhanced by the 3D effect[67], corresponding to a clear shortening of the ignition delay time in FIG. **25**(b). Furthermore, with this reaction wave, compared to ISBI, much higher temperatures are achieved inside the whole gas bubble, as shown in FIG. **25**(c, d). By comparing the TBD of RSBI and ISBI in FIG. **24**(b), we can find that the reacting effect has significantly suppressed the bubble expansion in the transverse direction, especially in the later stage. This is demonstrated by comparing the bubble shape (volume rendering of the mass fraction of Xe) of RSBI and ISBI in FIG. **26**, where we also observe that the bubble expansion in the streamwise direction is reduced as well by the reaction wave. Concerning the mixing between the bubble and the ambient gas, FIG. **24**(a) indicates that the reaction essentially diminishes the mixing of SBI after the shock passage. Consequently, in FIG. **27**, the vorticity distribution of the bubble at 76.5 μs, 200.5 μs, and 499.5 μs in RSBI is much higher than that in ISBI.

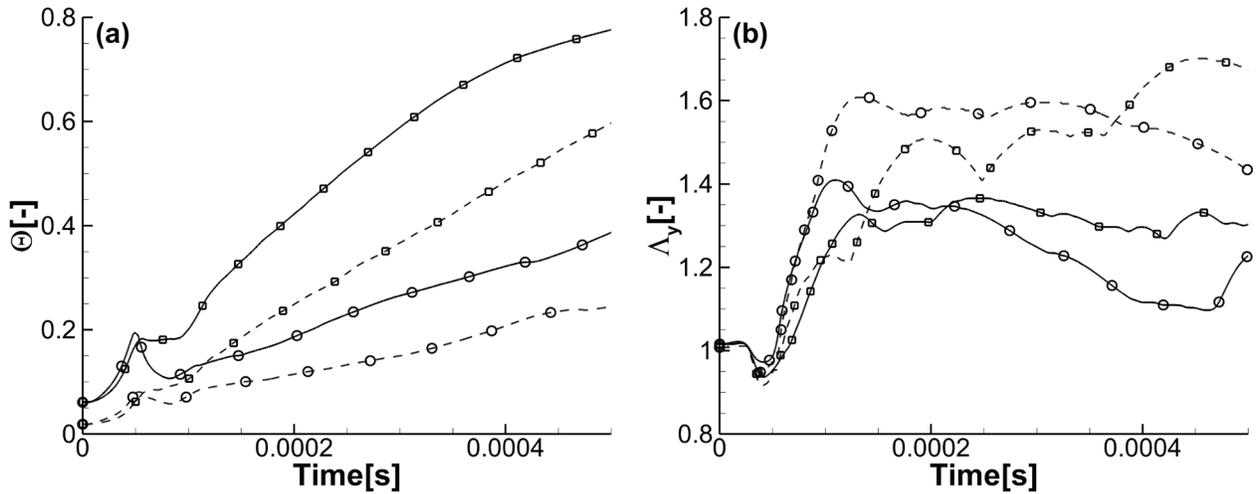

FIG. 24. The comparison of (a) MMF and (b) TBD between the 2D and 3D SBI simulations. The solid and dashed lines represent the 3D and 2D, receptively. The square and circle symbols represent ISBI and RSBI problems, receptively.

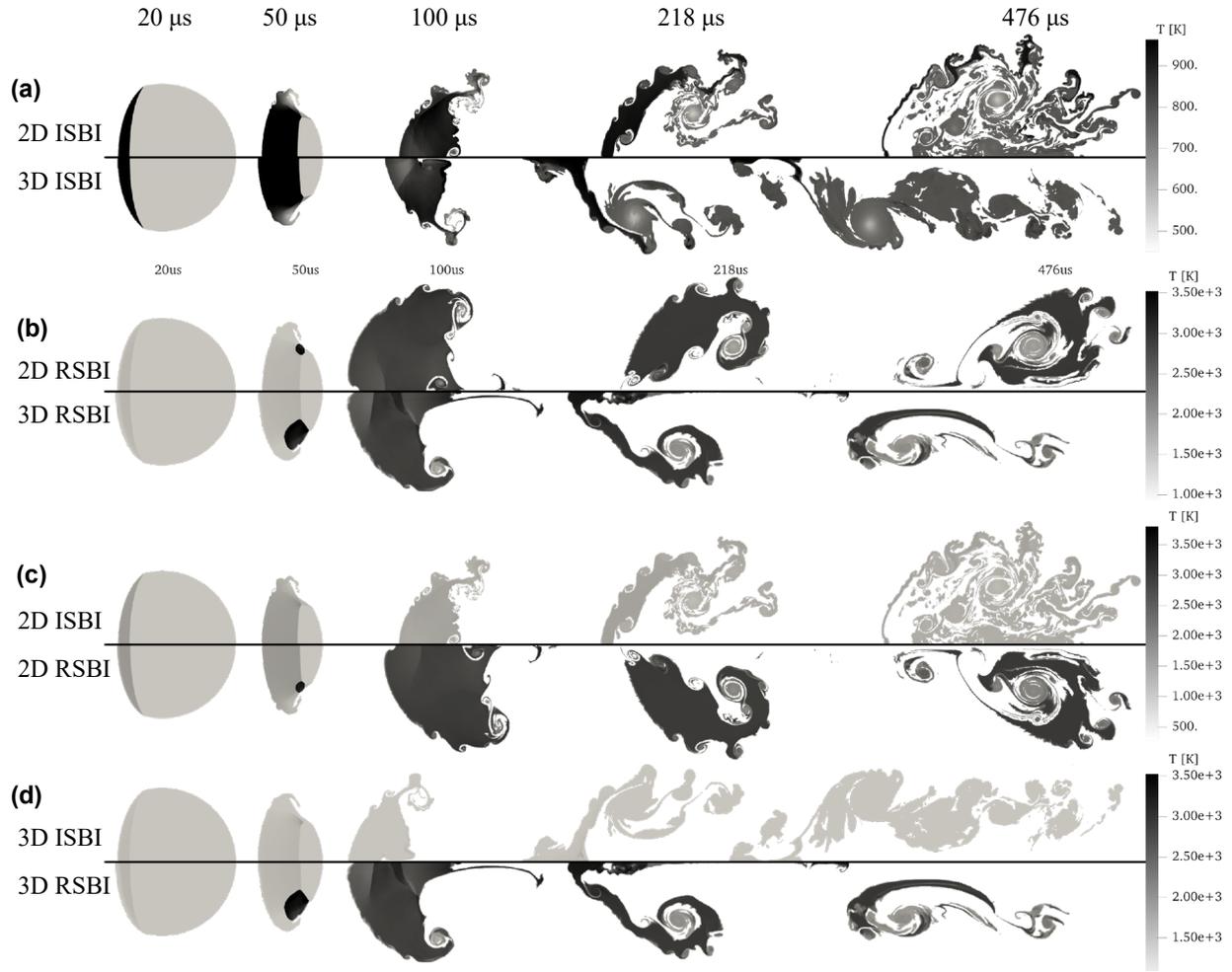

FIG. 25. The contours of the temperature field for 2D and 3D SBI simulations at 5 different time instants: the comparison between 2D and 3D results for (a) the inert case and (b) the reacting case, the comparison between the inert and reacting cases in (c) 2D and (d) 3D.

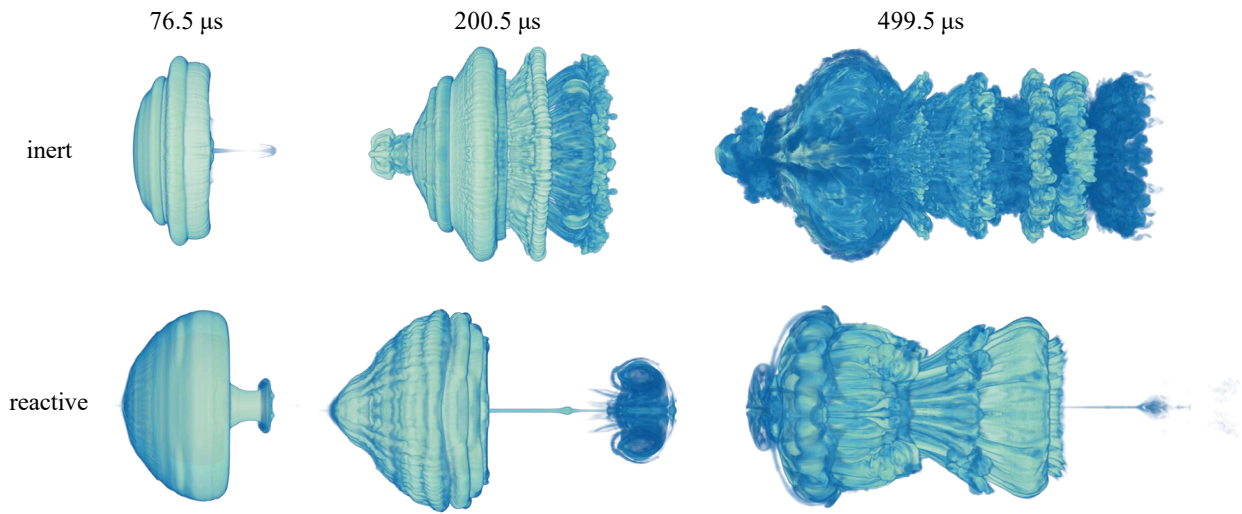

FIG. 26. The bubble shape represented by volume rendering of the mass fraction of Xe for 3D SBI simulations.

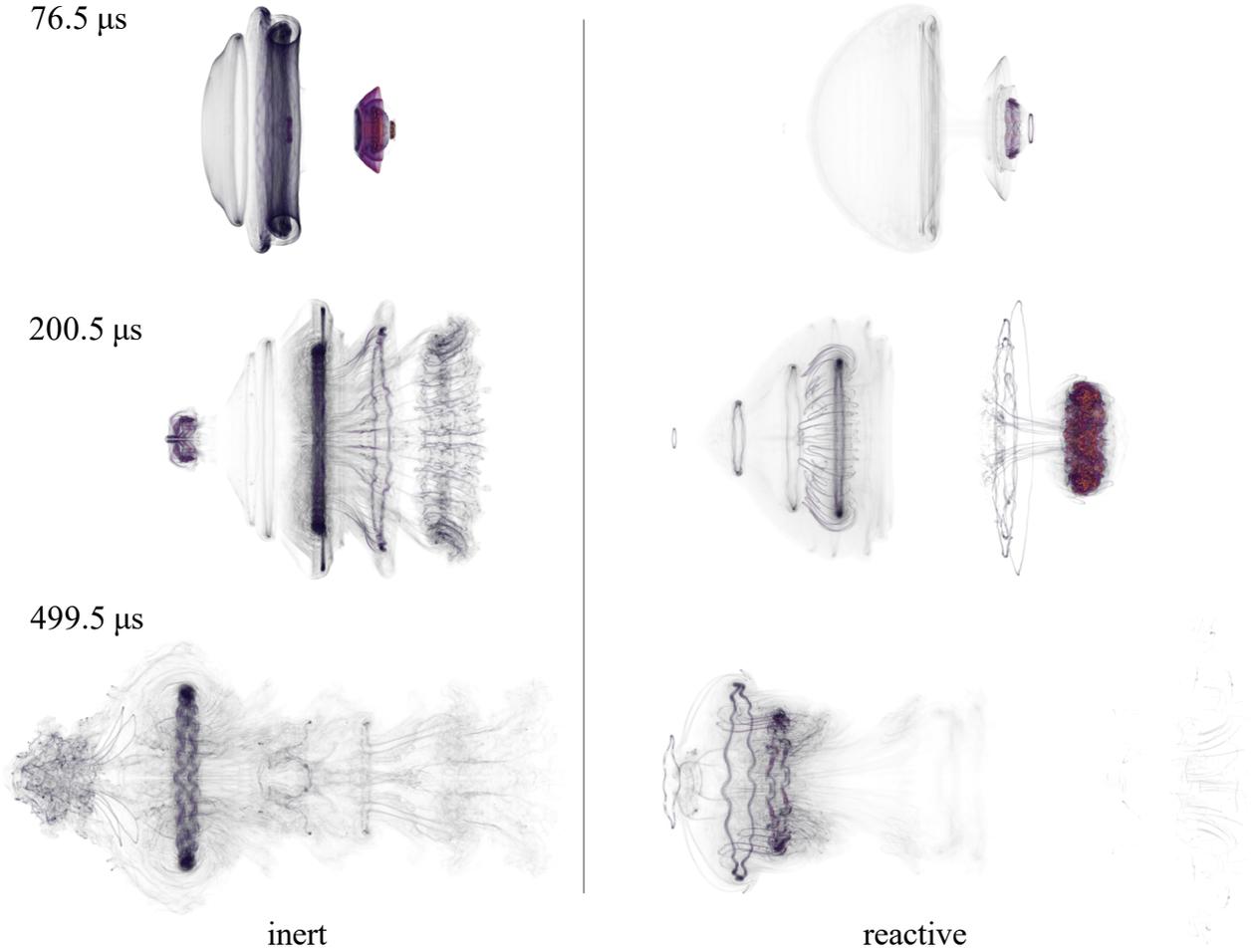

FIG. 27. The volume rendering of vorticity for the 3D SBI simulations.

## VII. CONCLUSION

This work introduces an open-source heterogeneous solver, XFluids, for high-resolution and multi-GPU parallel simulations of compressible viscous reacting multicomponent flows on structured meshes. Based purely on SYCL, XFluids is characterized by a unified cross-architecture feature and is capable of communicating across multiple GPUs with GPU-enabled MPI. To accelerate the most time-consuming parts of XFluids, such as the reaction integral and convection reconstruction, numerous optimization strategies are applied, e.g. the adaptive range assignment, partial eigensystem reconstruction, and fitting transport coefficients. Multiple validation cases are performed to demonstrate that XFluids can be correctly executed in different GPUs of NVIDIA, AMD, and Intel. Compared to native GPU programming models like CUDA and HIP, XFluids shows competitive performance on different devices, with a weak-scaling parallel efficiency of up to 97% on 1024 AMD GPUs. Finally, we apply XFluids to simulate both the inert and reactive multicomponent SBI problems with 500 million high-resolution meshes, demonstrating its accuracy, robustness, and efficiency in solving large-scale complex fluid dynamics. In conclusion, XFluids have the following key features for reacting flow simulations: a) it has high performance for high-order discretization of convective-viscous terms and α-QSS time integration of stiff reactions; b) it can be easily offloaded to a range of heterogeneous devices without translating any source code; c) it achieves high efficiency and high parallel scalability on multi-GPU platforms.

**DECLARATION OF COMPETING INTERST**

The authors declare that they have no known competing financial interests or personal relationships that could have appeared to influence the work reported in this paper.

**ACKNOWLEDGMENTS**

This work was supported by the National Natural Science Foundation of China (Grant No. 11902271), the Guanghe foundation (No. ghfund202302016412), and the 111 project on Aircraft Complex Flows and the Control (Grant No. B17037). We acknowledge Yixuan Lian for useful discussion about viscous fluxes discretization methods, and Renfei Zhang for some tests and data processing of XFluids and thank the AdaptiveCpp community.

**DATA AVAILABILITY**

The data that support the findings of this study are available from the corresponding author upon reasonable request.

# APPENDIX A: A simple SYCL code

Main SYCL code:

```cpp
#include <stdio.h>
#include <string.h>
#include <sycl/sycl.hpp>

struct ProcPara{int *a, *b, *c;}, Block { int a, b, c;};
// Device Code
int addDevice(int a, Block bl)
{
    sycl::ext::oneapi::experimental::printf("Device function called && bl.c= %d \n", bl.c);
    return a * bl.c;
}
// Kernels
extern SYCL_EXTERNAL void addKernel(ProcPara *d_para, Block bl, sycl::nd_item<3> item_ct1)
{
    int i = item_ct1.get_local_id(2);
    d_para->c[i] = d_para->a[i] * bl.a + d_para->b[i] * bl.b;
    d_para->c[i] = addDevice(d_para->c[i], bl);
    sycl::ext::oneapi::experimental::printf("Kernel function called && d_para->c[i]= %d \n", d_para->c[i]);
}
void addBlock(ProcPara *d_para, Block bl, sycl::queue &q)
{
    // Throw a kernel expression parallelism to the GPU.
    q.submit([&](sycl::handler &h){
        sycl::stream stream_ct1(64 * 1024, 80, h);//for Output in kernel and device function
        h.parallel_for(sycl::nd_range<3>(sycl::range<3>(1, 1, bl.c), sycl::range<3>(1, 1, bl.c)),
            [=](sycl::nd_item<3> item_ct1) {
                addKernel(d_para, bl, item_ct1 );
            }); }).wait();
}
int main()
{
    Block bl;
    const int arraySize = 5;
    ProcPara *h_para, *d_para;
    // Select Device
    auto device = sycl::platform::get_platforms()[DEV].get_devices()[0];
    sycl::queue q(device);
    // Allocate Memory
    h_para = static_cast<ProcPara *>(sycl::malloc_host(sizeof(ProcPara), q));
    (h_para)->a = static_cast<int *>(sycl::malloc_host(arraySize * sizeof(int), q));
    (h_para)->b = static_cast<int *>(sycl::malloc_host(arraySize * sizeof(int), q));
    (h_para)->c = static_cast<int *>(sycl::malloc_host(arraySize * sizeof(int), q));
    d_para = static_cast<ProcPara *>(sycl::malloc_device(sizeof(ProcPara), q));
    // Initialization
    bl.a = 2,
    bl.b = 3, bl.c = arraySize;
    for (size_t i = 0; i < arraySize; i++)
        h_para->a[i] = i, h_para->b[i] = 10 * i, h_para->c[i] = 0;
    // Migrate Memory from Host to Device
    q.memcpy(d_para, h_para, sizeof(ProcPara)).wait();
    // Throw parallelism to Device
    addBlock(d_para, bl, q);
    // Migrate Memory from Device to host
    q.memcpy(h_para, d_para, sizeof(ProcPara)).wait();
    return 0;
}
```

Compiling options:

```makefile
build-host-backend:
	clang++ -DDEV=1 -fsycl -O0 ./main.cpp -o HOST-test

build-nvidia-backend:
	clang++ -DDEV=2 -fsycl -O0 -fsycl-targets=nvptx64-nvidia-cuda -Xsycl-target-backend --cuda-gpu-arch=sm_75 ./main.cpp -o NVIDIA-test

build-amd-backend:
	clang++ -DDEV=2 -fsycl -O0 -fsycl-targets=amdgcn-amd-amdhsa -Xsycl-target-backend --offload-arch=gfx906 ./main.cpp -o AMD-test

clean:
	rm -rf ./*-test
```